\documentclass[review,12pt,3p]{elsarticle}

\usepackage{tikz}
\usepackage{amsmath} 
\usepackage{amssymb}
\usetikzlibrary{shapes,arrows,chains,intersections}
\usepackage{lineno}
\modulolinenumbers[5]

\usepackage{subfigure}
\usepackage{float}
\usepackage{textcomp} 
\usepackage{url}

\graphicspath{{./pdf/}}

\usepackage{natbib}
\usepackage{multicol} 
\usepackage{nomencl}
\usepackage{framed} 
\makenomenclature
\renewcommand*\nompreamble{\begin{multicols}{2}}
\renewcommand*\nompostamble{\end{multicols}}

\usepackage{multirow}

\journal{Journal of Computational Physics}

\newcommand{\vect}[1]{\boldsymbol{#1}}
\newtheorem{remark}{Remark}
\begin{document}


\begin{frontmatter}

\title{A sufficient condition for free-stream preserving in the nonlinear conservative finite difference schemes on curvilinear grids}

\author{Hongmin Su}
\ead{hongminsu@mail.nwpu.edu.cn}

\author{Jinsheng Cai\corref{cor1}}
\ead{caijsh@nwpu.edu.cn}

\author{Kun Qu\corref{cor3}}
\ead{kunqu@nwpu.edu.cn}

\author{Shucheng Pan\corref{cor2}}
\ead{shucheng.pan@nwpu.edu.cn}

\cortext[cor3]{Corresponding author}

\address{Department of Fluid Mechanics, School of Aeronautics, Northwestern Polytechnical University, Xi'an 710072, China}

\begin{abstract}
In simulations of compressible flows, the conservative finite difference method (FDM) based on the nonlinear upwind schemes, e.g. WENO5, might violate free-stream preserving (FP), due to the loss of the geometric conservation law (GCL) identity when applied on the curvilinear grids.
Although some techniques on FP have been proposed previously, no general rule is given for this issue.
In this paper, by rearranging the upwind dissipation of the nonlinear schemes as a combination of sub-stencil reconstructions (taking WENO5 as an example), it can be proved that the upwind dissipation diminishes under the uniform flow condition if the metrics yield an identical value under the same schemes with these reconstructions, making the free-stream condition be preserved.
According to this sufficient condition, the novel FP metrics are constructed for WENO5 and WENO7. By this means the original forms of these WENO schemes can be kept. In addition, the accuracy of these schemes can be retained as well with a simple accuracy compensation by replacing the central part fluxes with a high-order one.
Various validations indicate that the present FP schemes retain the great capability to  resolve the smooth regions accurately and capture the discontinuities robustly.

\end{abstract}

\begin{keyword}
Conservative finite difference method \sep geometric conservative law \sep free-stream preserving  \sep high-order schemes \sep WENO scheme
\end{keyword}

\end{frontmatter}

\section{Introduction}
The GCL~\cite{thomas1979geometric}, including volume conservation law (VCL) and surface conservation law (SCL)~\cite{zhang1993discrete}, is very important in computational fluid dynamics, especially for the high-accuracy simulations.
Unlike finite volume method (FVM) and finite element method (FEM), the conservative FDM binds the physical quantities and geometric metrics together during the process of flux reconstruction in the computational space such that the GCL is not easy to be satisfied due to the discretization errors of grid metrics for the upwind dissipation.
The violation of the GCL will yield large errors, oscillations and instabilities in simulations~\cite{thomas1979geometric,VISBAL2002155,NONOMURA2010197}, and even lead to the non-conservation of the governing equations~\cite{Pulliam1978,zhang1993discrete}.
On stationary grids, the VCL identity is satisfied naturally but the SCL identity might not be satisfied, which yields the freestream preserving(FP) problems.

Many achievements have been put forward to maintain the FP identity, especially for the low order schemes, as concluded in Ref.~\cite{DENG20111100,nonomura_new_2015}.
In the high-accuracy FDM, the symmetry conservative metric method(SCMM)~\cite{vinokur2002extension}, developed from the conservative metrics in Ref.~\cite{thomas1979geometric}, have been an efficient technique to fulfill this identity under the sufficient condition of evaluating the derivatives of the grid metrics and the convection fluxes with the unique scheme given by Deng et al.~\cite{DENG201390} and Abe et al.~\cite{abe2014geometric}, which is a milestone in this area. And the cell-center version of the SCMM is also developed by Liao et al. \cite{Fei2015Extending1,Fei2015Extending2}.
However, for the upwind schemes which are very important in the numerical simulations of compressible flows, it is not easy to be achieved due to the inconsistent differential operators applied for the grid metrics and fluxes~\cite{DENG20111100,ABE201314}.
At present, there are mainly two ideas to deal with the free-stream conservation problem for high-order upwind schemes. The first one is to consider the independent interpolation for flow variables and metrics, such as WCNS~\cite{DENG200022,DENG20111100,DENG201390}, alternative finite-difference form of WENO (AWENO)~\cite{Jiang2013An, jiang2018, Yu2020}, etc..
In these schemes, the dissipation is handled by the finite volume method and then obtain their derivatives by central schemes.
The second method is to separate the central part from the standard upwind schemes and employ the high-order central schemes to it while the dissipative part is handled by the finite-volume-like methods. For examples, Nonomura et al.~\cite{nonomura_new_2015} and Zhu et al.~\cite{zhu_free-stream_2019} apply the frozen metrics for the entire stencils or the local difference form partially, and Li et al.~\cite{LiFurther} replaces the transformed conservative variables with the original one.
Recently, Su et al. \cite{Su2020FP} suggest reformulating the upwind dissipation by subtracting a reference cell-face flow state from each cell-center flux in the local stencil, which gives a simple and efficient strategy on FP.
However, these schemes either bring more or less additional complications or result in large approximation errors due to the metrics freezing.
More than this, in conservative FDM, unlike the linear schemes which the SCMM under the sufficient condition of Deng et al.~\cite{DENG201390} and Abe et al.~\cite{abe2014geometric} can be applied in directly to achieve FP identity, the nonlinear upwind schemes have no general rule to achieve this identity, which results in many difficulties in practical applications.

In this study, we carry out a further investigate on the dissipation of 5th-order WENO scheme in conservative FDM and conclude a common rule for the previous FP methods. With this, we give a supplementary sufficient condition for the nonlinear schemes to maintain the FP identity. After that, together with the previous sufficient condition of Deng et al.~\cite{DENG201390} and Abe et al.~\cite{abe2014geometric} which can be applied to the central schemes, a simple, efficient and high-order FP technique for WENO is proposed. The present method employs the new FP metrics on the WENO schemes straightforwardly with a simple accuracy compensation, which possesses at least three advantages: (1) maintains the theoretical convergence orders of the upwind schemes; (2) avoids modifying the original forms of these schemes; (3) provides convenience for the extension to other upwind schemes. The outline is organized as follows. In sec. 2, we introduce the governing equations, the upwind schemes, the SCL identity and the previous FP method~\cite{nonomura_new_2015,zhu_free-stream_2019,Su2020FP} in conservative FDM. Then, a supplementary sufficient condition for WENO schemes to maintain the FP identity is given and the new FP metrics are proposed under this condition in Sec. 3. Next, various validations and numerical examples to evaluate the accuracy and robustness of the present FP method are given in Sec. 4. Finally, a brief conclusion is available in Sec. 5.

\section{Governing equations and metrics on stationary curvilinear coordinates}
\subsection{Navier-Stokes equations}
The compressible Navier-Stokes equations on curvilinear grids are formulated by
\begin{equation} \label{eq:NS}
   \begin{aligned}
\dfrac{\partial \vect{\tilde{Q}}}{\partial t}&+\dfrac{\partial \vect{\tilde{F}}}{\partial \xi} + \dfrac{\partial \vect{\tilde{G}}}{\partial \eta} +\dfrac{\partial \vect{\tilde{H}}}{\partial \zeta} -\dfrac{\partial \vect{\tilde{F_v}}}{\partial \xi} -\dfrac{\partial \vect{\tilde{G_v}}}{\partial \eta} -\dfrac{\partial \vect{\tilde{H_v}}}{\partial \zeta} =\vect{0},
\end{aligned}
\end{equation}
with an equation of state for ideal gas
\begin{equation} \label{eq:EOS}
\begin{aligned}
   p&=\left( \gamma-1\right) \rho e,\\
   \gamma&=1.4.
\end{aligned}
\end{equation}
Here, $t$ stands for the physical time, and $\xi,\eta,\zeta$, $J$ are the transformed coordinates and Jacobian on a uniform computational domain. Besides,
\begin{equation} \label{eq:Q}
   \boldsymbol{\tilde{Q}}=\dfrac{ \boldsymbol{Q} }{J},
\end{equation}
\begin{equation} \label{eq:FGH}
\begin{aligned}
\boldsymbol{\tilde{F}}&=\dfrac{\xi_x \boldsymbol{F}+\xi_y \boldsymbol{G}+\xi_z   \boldsymbol{H}}{J}, \\
   \boldsymbol{\tilde{G}}&=\dfrac{\eta_x \boldsymbol{F}+\eta_y \boldsymbol{G}+\eta_z \boldsymbol{H}}{J},\\
   \boldsymbol{\tilde{H}}&=\dfrac{\zeta_x \boldsymbol{F}+\zeta_y \boldsymbol{G}+\zeta_z \boldsymbol{H}}{J},
\end{aligned}
\end{equation}
\begin{equation} \label{eq:FGH_v}
\begin{aligned}
   \boldsymbol{\tilde{F_v}}&=\dfrac{\xi_x \boldsymbol{F_v}+\xi_y \boldsymbol{G_v}+\xi_z \boldsymbol{H_v}}{J},\\
   \boldsymbol{\tilde{G_v}}&=\dfrac{\eta_x \boldsymbol{F_v}+\eta_y \boldsymbol{G_v}+\eta_z \boldsymbol{H_v}}{J},\\
\boldsymbol{\tilde{H_v}}&=\dfrac{\zeta_x \boldsymbol{F_v}+\zeta_y \boldsymbol{G_v}+\zeta_z \boldsymbol{H_v}}{J}.
\end{aligned}
\end{equation}
The fluxes $\boldsymbol{F}$, $\boldsymbol{G}$, $\boldsymbol{H}$ and $\boldsymbol{F_v}$, $\boldsymbol{G_v}$, $\boldsymbol{H_v}$ represent the inviscid and viscous flux vectors in x, y and z direction, respectively. In details,
\begin{equation} \label{eq:Q}
  \boldsymbol{Q}=\left(
        \begin{matrix}
          \rho & \rho u_1  &  \rho u_2   &  \rho u_3    &   \rho E
        \end{matrix}
        \right)^T,
\end{equation}
\begin{equation} \label{eq:F}
\boldsymbol{F}=\left(
\begin{matrix}
          \rho u_1 & \rho u_1 u_1+p & \rho u_2 u_1 & \rho u_3 u_1  & (\rho E+p) u_1
\end{matrix}
        \right)^T,
\end{equation}
\begin{equation} \label{eq:G}
\boldsymbol{G}=\left(
\begin{matrix}
          \rho u_2 & \rho u_1 u_2 & \rho u_2 u_2+p  &\rho u_3 u_2 & (\rho E+p) u_2
\end{matrix}
        \right)^T,
\end{equation}
\begin{equation} \label{eq:H}
\boldsymbol{H}=\left(
\begin{matrix}
          \rho u_3 & \rho u_1 u_3 & \rho u_2 u_3 & \rho u_3 u_3+p  & (\rho E+p) u_3
\end{matrix}
        \right)^T,
\end{equation}
\begin{equation} \label{eq:NSv1}
\vect{F_v}=\left(
        \begin{matrix}
           0& \tau_{11} & \tau_{12} & \tau_{13}& u_i\tau_{i1}-\dot{q}_1
        \end{matrix}
        \right)^T
\end{equation}
\begin{equation} \label{eq:NSv2}
\vect{G_v}=\left(
        \begin{matrix}
           0& \tau_{21} & \tau_{22} & \tau_{23}& u_i\tau_{i2}-\dot{q}_2
        \end{matrix}
        \right)^T
\end{equation}
\begin{equation} \label{eq:NSv3}
\vect{H_v}=\left(
        \begin{matrix}
           0& \tau_{31} & \tau_{32} & \tau_{33}& u_i\tau_{i3}-\dot{q}_3
        \end{matrix}
        \right)^T
\end{equation}
$u_1$, $u_2$, $u_3$ are the velocity components. $\rho$, $p$ and $E$ are the density, pressure and the total specific energy. $\tau_{ij}$ denotes the shear stress tensor
\begin{equation} \label{eq:stress}
   \tau_{ij}=2\mu \left(S_{ij}-\delta_{ij}\dfrac{S_{kk}}{3} \right),
\end{equation}
\begin{equation} \label{eq:strain}
   S_{ij}=\dfrac{1}{2} \left(\dfrac{\partial u_i}{\partial x_j}+\dfrac{\partial u_j}{\partial x_i} \right),
\end{equation}
and $\dot{q}_i$ is the heat flux in direction i
\begin{equation}
  \dot{q}_{i}=-\kappa \dfrac{\partial T}{\partial x_i},
\end{equation}
where $\mu$ and $\kappa$ is the shear viscosity and thermal conductivity.

\subsection{Discretization methods}
 The conservative FDM~\cite{Shu1988Efficient,Shu1989Efficient,Merriman2003Understanding} reconstructs the high-order consistent numerical flux at each cell-face. Without loss of generality, the fluxes $ \boldsymbol{\tilde{F}}_{i}$ in $\xi$ direction indexed by $i$ is regarded as an average of a primitive function $\boldsymbol{\hat{H}}(\xi)$
\begin{equation} \label{eq:avr}
\boldsymbol{\tilde{F}}_{i}=\dfrac{1}{\Delta \xi}\int_{i-1/2}^{i+1/2} \boldsymbol{\hat{H}}(\xi)d\xi
\end{equation}
Then we can exactly obtain the derivative of $\boldsymbol{\tilde{F}}_{i}$,
\begin{equation} \label{eq:derivative}
\left( \dfrac{\partial \boldsymbol{\tilde{F}}}{\partial \xi} \right)_i=\dfrac{\boldsymbol{\hat{H}}_{i+1/2}-\boldsymbol{\hat{H}}_{i-1/2}}{\Delta \xi}.
\end{equation}
Therefore, the derivative of the convective fluxes can be approximated by the reconstructed cell-face fluxes $\boldsymbol{\tilde{F}}_{i+1/2}$
\begin{equation} \label{eq:aderivative}
\left( \dfrac{\partial \boldsymbol{\tilde{F}}}{\partial \xi} \right)_i=\dfrac{ \boldsymbol{\tilde{F}}_{i+1/2}- \boldsymbol{\tilde{F}} _{i-1/2}}{\Delta \xi}+\boldsymbol{O}(\Delta \xi^{2k-1}),
\end{equation}
where $\boldsymbol{\tilde{F}}_{i+1/2}$ is the approximation of the primitive function value at cell-face $\boldsymbol{\hat{H}}_{i+1/2}$, which can be reconstructed by upwind schemes from the cell fluxes $\boldsymbol{\tilde{F}}_{i-k+1}$, $\cdots$, $\boldsymbol{\tilde{F}}_{i+k-1}$ to keep the $(2k-1)$th-order accuracy, such as WENO scheme.

\subsubsection{The characteristic-wise WENO scheme}\label{sec:WENO}
For the purpose of improving the robustness of the simulations, the fluxes and conservative variables are transformed into the characteristic space and then a flux vector splitting scheme, such as local Lax-Friedrichs splitting, is applied,
\begin{equation}\label{eq:LF_splitting}
 \boldsymbol{ \tilde{F} }_m^{\pm}= \dfrac{1}{2} \boldsymbol{L}_{i+1/2} \cdot \left( \boldsymbol{\tilde{F}}_m \pm \boldsymbol{\Lambda}_{i+1/2} \boldsymbol{\tilde{Q}}_m \right), m=i-k+1,\cdots,i+k,
\end{equation}
where
$\boldsymbol{\Lambda}_{i+1/2}$ is the diagonal matrix composed by $\lambda^s=max\left( |\lambda_m^s| \right)$, which is the maximum s-th eigenvalue of $\boldsymbol{A}_{i+1/2}=\left( \partial \boldsymbol{\tilde{F}} /\partial \boldsymbol{\tilde{Q}} \right)_{i+1/2}$ cross the stencil. And $\boldsymbol{L}_{i+1/2}$, $\boldsymbol{R}_{i+1/2}$ are the matrix of left and right eigenvectors of $\boldsymbol{A}_{i+1/2}$, respectively. 
After obtaining $\tilde{\mathbf{F}}^{\pm}_{i+1/2}$ by WENO reconstruction, the cell-face fluxes are given by
 \begin{equation}\label{eq:Fp+Fm}
 \boldsymbol{ \tilde{F} }_{i+1/2}= \boldsymbol{R}_{i+1/2} \cdot \left( \boldsymbol{ \tilde{F} }_{i+1/2}^{+}+\boldsymbol{ \tilde{F} }_{i+1/2}^{-} \right).
\end{equation}

In the nonsmooth region, the cell-face fluxes $\boldsymbol{ \tilde{F} }_{i+1/2}^{\pm}$ can be reconstructed by the nonlinear schemes. Here we choose WENO5~\cite{jiang_efficient_1996} to obtain these fluxes by
\begin{equation}\label{eq:weno_formulation}
\widetilde{f}_{i+1/2}^{\pm}=\sum\limits_{k=0}^{2}\omega_k^{\pm} q_k^{\pm},
\end{equation}
where $\widetilde{f}^{\pm}$ denotes one of the component of $ \boldsymbol{ \tilde{F} }^{\pm}$. Taking the positive fluxes as an example, the three 3rd-order approximations for the different sub-stencils are formulated as
\begin{equation}\label{eq:substencil}
\begin{aligned}
q_0^{+}&= \dfrac{1}{3}\widetilde{f}_{i-2}^{+}
          -\dfrac{7}{6}\widetilde{f}_{i-1}^{+}
          +\dfrac{7}{6}\widetilde{f}_{i}^{+},\\
q_1^{+}&=-\dfrac{1}{6}\widetilde{f}_{i-1}^{+}
          +\dfrac{5}{6}\widetilde{f}_{i}^{+}
          +\dfrac{1}{3}\widetilde{f}_{i+1}^{+},\\
q_2^{+}&= \dfrac{1}{3}\widetilde{f}_{i}^{+}
          +\dfrac{5}{6}\widetilde{f}_{i+1}^{+}
          -\dfrac{1}{6}\widetilde{f}_{i+2}^{+}.
\end{aligned}
\end{equation}
The nonlinear weight $\omega_k^{+}$ in Eq.\eqref{eq:weno_formulation} is evaluated by
\begin{equation}
\omega_k^{+}=\dfrac{C_k}{\left( \beta_k^{+}+\epsilon \right)^n} \bigg/ \sum\limits_{r=0}^{2}\dfrac{C_r}{\left( \beta_k^{+}+\epsilon \right)^n},
\end{equation}
and the smooth indicators $\beta_k^{+}$ are determined by
\begin{equation}
\begin{aligned}
\beta_0^{+}&= \dfrac{1}{4} \left(  \widetilde{f}_{i-2}^{+}
                                 -4\widetilde{f}_{i-1}^{+}
                                 +3\widetilde{f}_{i}^{+} \right)^2
            +\dfrac{13}{12}\left(  \widetilde{f}_{i-2}^{+}
                                 -2\widetilde{f}_{i-1}^{+}
                               +\widetilde{f}_{i}^{+} \right)^2,\\
\beta_1^{+}&= \dfrac{1}{4} \left( -\widetilde{f}_{i-1}^{+}
                               +\widetilde{f}_{i+1}^{+} \right)^2
            +\dfrac{13}{12}\left(  \widetilde{f}_{i-1}^{+}
                                 -2\widetilde{f}_{i}^{+}
                              +\widetilde{f}_{i+1}^{+} \right)^2,\\
\beta_2^{+}&= \dfrac{1}{4} \left(-3\widetilde{f}_{i}^{+}
                                 +4\widetilde{f}_{i+1}^{+}
                                -\widetilde{f}_{i+2}^{+} \right)^2
            +\dfrac{13}{12}\left(  \widetilde{f}_{i}^{+}
                                 -2\widetilde{f}_{i+1}^{+}
                              +\widetilde{f}_{i+2}^{+} \right)^2,\\
\end{aligned}
\end{equation}
where $\epsilon=1.0\times10^{-6}$ and $n=2$. Moreover, $C_0=1/10$, $C_1=3/5$, $C_2=3/10$ are the optimal weights which recover WENO5 to the 5th-order linear upwind scheme (LU5) given by
 \begin{equation}\label{eq:linear_upwind}
 \begin{aligned}
\boldsymbol{ \tilde{F} }_{i+1/2}^{+}&=\dfrac{1}{60}\left( 2\boldsymbol{ \tilde{F} }_{i-2}^{+}-13\boldsymbol{ \tilde{F} }_{i-1}^{+}+47\boldsymbol{ \tilde{F} }_{i}^{+}+27\boldsymbol{ \tilde{F} }_{i+1}^{+}-3\boldsymbol{ \tilde{F} }_{i+2}^{+} \right), \\
\boldsymbol{ \tilde{F} }_{i+1/2}^{-}&=\dfrac{1}{60}\left( -3\boldsymbol{ \tilde{F} }_{i-1}^{-}+27\boldsymbol{ \tilde{F} }_{i}^{-}+47\boldsymbol{ \tilde{F} }_{i+1}^{-}-13\boldsymbol{ \tilde{F} }_{i+2}^{-}+2\boldsymbol{ \tilde{F} }_{i+3}^{-} \right),
\end{aligned}
\end{equation}
which can be only employed in smooth region. 

\subsection{SCL and SCMM}\label{sec:SCMM_SCL}
Imposing the free-stream condition to Navier-Stokes equations gives
\begin{equation} \label{eq:SCL}
\begin{aligned}
I_x= \left( \dfrac{ \xi_x}{J} \right)_{\xi}
    +\left( \dfrac{ \eta_x}{J} \right)_{\eta}
    +\left( \dfrac{ \zeta_x}{J}\right)_{\zeta}=0, \\
I_y= \left( \dfrac{ \xi_y}{J} \right)_{\xi}
    +\left( \dfrac{ \eta_y}{J} \right)_{\eta}
    +\left( \dfrac{ \zeta_y}{J}\right)_{\zeta}=0,\\
I_z= \left( \dfrac{ \xi_z}{J} \right)_{\xi}
    +\left( \dfrac{ \eta_z}{J} \right)_{\eta}
    +\left( \dfrac{ \zeta_z}{J}\right)_{\zeta}=0.
\end{aligned}
\end{equation}
These equations are regarded as the SCL by Zhang et al.~\cite{zhang1993discrete} because they delineate the consistence of vectorized computational cell surfaces in finite volume method~\cite{VINOKUR19891}. However, the discretized errors induced by finite difference schemes can easily destroy this identity such that the free-stream condition can not be preserved for the simulations on curvilinear grids, which refers to the FP problem.
To satisfy the FP identity, the SCMM under the sufficient condition in Ref.~\cite{DENG201390} and \cite{abe2014geometric}
is the most efficient method in high order finite difference simulations so far, such as WCNS~\cite{DENG201390}. The symmetry conservative metrics are expressed as
\begin{equation} \label{eq:JBabc}
\begin{aligned}
  \dfrac{\xi_x}{J}=\dfrac{1}{2} \left[
        \left( y_\eta z \right)_\zeta
     -  \left( y_\zeta z \right)_\eta
     +  \left( yz_\zeta\right)_\eta
     -  \left( yz_\eta \right)_\zeta
     \right],  \\
  \dfrac{\xi_y}{J}=\dfrac{1}{2} \left[
        \left( xz_\eta \right)_\zeta
     -  \left( xz_\zeta \right)_\eta
     +  \left( x_\zeta z\right)_\eta
     -  \left( x_\eta z \right)_\zeta
     \right],  \\
  \dfrac{\xi_z}{J}=\dfrac{1}{2} \left[
        \left( x_\eta y \right)_\zeta
     -  \left( x_\zeta y \right)_\eta
     +  \left( xy_\zeta\right)_\eta
     -  \left( xy_\eta \right)_\zeta
     \right],  \\
  \dfrac{\eta_x}{J}=\dfrac{1}{2} \left[
        \left( y_\zeta z \right)_\xi
     -  \left( y_\xi z \right)_\zeta
     +  \left( yz_\xi  \right)_\zeta
     -  \left( yz_\zeta \right)_\xi
     \right],  \\
  \dfrac{\eta_y}{J}=\dfrac{1}{2} \left[
        \left( xz_\zeta\right)_\xi
     -  \left( xz_\xi  \right)_\zeta
     +  \left( x_\xi z  \right)_\zeta
     -  \left( x_\zeta z \right)_\xi
     \right],  \\
  \dfrac{\eta_z}{J}=\dfrac{1}{2} \left[
        \left( x_\zeta y\right)_\xi
     -  \left( x_\xi y \right)_\zeta
     +  \left( xy_\xi  \right)_\zeta
     -  \left( xy_\zeta \right)_\xi
     \right],  \\
  \dfrac{\zeta_x}{J}=\dfrac{1}{2} \left[
        \left(y_\xi z\right)_\eta
     -  \left(y_\eta z \right)_\xi
     +  \left( yz_\eta  \right)_\xi
     -  \left( yz_\xi \right)_\eta
     \right],  \\
  \dfrac{\zeta_y}{J}=\dfrac{1}{2} \left[
        \left(xz_\xi \right)_\eta
     -  \left(xz_\eta \right)_\xi
     +  \left( x_\eta z  \right)_\xi
     -  \left( x_\xi z \right)_\eta
     \right],  \\
  \dfrac{\zeta_z}{J}=\dfrac{1}{2} \left[
        \left(x_\xi y \right)_\eta
     -  \left(x_\eta y \right)_\xi
     +  \left( xy_\eta  \right)_\xi
     -  \left( xy_\xi  \right)_\eta
     \right],  \\
\end{aligned}
\end{equation}
and
\begin{equation} \label{eq:JB}
  \dfrac{1}{J}=\dfrac{1}{3} \left[
        \left( x \dfrac{\xi_x}{J}+y \dfrac{\xi_y}{J}+z \dfrac{\xi_z}{J}  \right)_\xi
       +\left( x \dfrac{\eta_x}{J}+y \dfrac{\eta_y}{J}+z \dfrac{\eta_z}{J}  \right)_\eta
       +\left( x \dfrac{\zeta_x}{J}+y \dfrac{\zeta_y}{J}+z \dfrac{\zeta_z}{J}  \right)_\zeta
     \right] .
\end{equation}
However, even if the SCMM is employed, the FP identity can not be maintained for the nonlinear upwind schemes in the conservative FDM as the metrics are calculated by the linear schemes, which conflicts with the sufficient condition mentioned above.

\subsection{Previous achievements in FP identity for WENO scheme}
First, to achieve FP identity for WENO5 scheme , two recent techniques~\cite{nonomura_new_2015,zhu_free-stream_2019} split the original scheme into the central part and the upwind dissipation part ($\boldsymbol{D}^{+}$ and $\boldsymbol{D}^{-}$),
\begin{equation}\label{eq:WENO_cen_dis}
\begin{aligned}
\boldsymbol{ \tilde{F} }_{i+1/2}
=&\underbrace{ \dfrac{1}{60} \left( \boldsymbol{\tilde{F}}_{i-2} -8 \boldsymbol{\tilde{F}}_{i-1}+37 \boldsymbol{\tilde{F}}_{i} +37\boldsymbol{\tilde{F}}_{i+1}-8 \boldsymbol{\tilde{F}}_{i+2}+\boldsymbol{\tilde{F}}_{i+3} \right) }_{\text{central part}}\\
&  \underbrace{ -\dfrac{1}{60} \sum\limits_s \boldsymbol{r}_{i+1/2}^s \left[ \left(20\omega_0^+-1 \right) \hat{f}_{i,1}^{s,+}-\left( 10\omega_0^+ +10\omega_1^+ -5 \right)\hat{f}_{i,2}^{s,+} +\hat{f}_{i,3}^{s,+}\right]}_{\text{upwind dissipation}, \boldsymbol{D}^+}\\
&  \underbrace{ +\dfrac{1}{60} \sum\limits_s \boldsymbol{r}_{i+1/2}^s \left[ \left(20\omega_0^- -1 \right) \hat{f}_{i,1}^{s,-}-\left( 10\omega_0^- +10\omega_1^- -5 \right)\hat{f}_{i,2}^{s,-} +\hat{f}_{i,3}^{s,-}\right]}_{\text{upwind dissipation}, \boldsymbol{D}^-},
\end{aligned}
\end{equation}
where
\begin{equation} \label{eq:WENO_cen_dis_sub1}
\begin{aligned}
\hat{f}_{i,r+1}^{s,+} =&\widetilde{f}_{i+r+1}^{s,+}-3\widetilde{f}_{i+r}^{s,+}+3\widetilde{f}_{i+r-1}^{s,+}-\widetilde{f}_{i+r-2}^{s,+} , \qquad r=0,1,2\\
=&\dfrac{1}{2} \boldsymbol{l}_{i+1/2}^s \left(  \boldsymbol{\tilde{F}}_{i+r+1}-3\boldsymbol{\tilde{F}}_{i+r}+3\boldsymbol{\tilde{F}}_{i+r-1}-\boldsymbol{\tilde{F}}_{i+r-2} \right) \\
&+\dfrac{1}{2}\lambda^s \boldsymbol{l}_{i+1/2}^s\left( \boldsymbol{\tilde{Q}}_{i+r+1}-3\boldsymbol{\tilde{Q}}_{i+r}+3\boldsymbol{\tilde{Q}}_{i+r-1}-\boldsymbol{\tilde{Q}}_{i+r-2} \right),
\end{aligned}
\end{equation}

\begin{equation}\label{eq:WENO_cen_dis_sub2}
\begin{aligned}
\hat{f}_{i,r+1}^{s,-}=&\widetilde{f}_{i-r+3}^{s,-}-3\widetilde{f}_{i-r+2}^{s,-}+3\widetilde{f}_{i-r+1}^{s,-}-\widetilde{f}_{i-r}^{s,-} , \qquad r=0,1,2\\
=&\dfrac{1}{2} \boldsymbol{l}_{i+1/2}^s \left(  \boldsymbol{\tilde{F}}_{i-r+3}-3\boldsymbol{\tilde{F}}_{i-r+2}+3\boldsymbol{\tilde{F}}_{i-r+1}-\boldsymbol{\tilde{F}}_{i-r} \right) \\
&-\dfrac{1}{2}\lambda^s \boldsymbol{l}_{i+1/2}^s\left( \boldsymbol{\tilde{Q}}_{i-r+3}-3\boldsymbol{\tilde{Q}}_{i-r+2}+3\boldsymbol{\tilde{Q}}_{i-r+1}-\boldsymbol{\tilde{Q}}_{i-r} \right).
\end{aligned}
\end{equation}
 According to Sec. \ref{sec:SCMM_SCL}, the central part in Eq. \eqref{eq:WENO_cen_dis} can successfully maintain the FP identity by adopting the SCMM with the sufficient condition of Deng et al.~\cite{DENG201390} and Abe et al.~\cite{abe2014geometric}. Specifically, the metrics are computed by
\begin{equation}\label{eq:metric_discret}
\begin{aligned}
x_{i+1/2}&=\dfrac{1}{60}\left( x_{i-2}-8x_{i-1}+37x_{i}+37x_{i+1}-8x_{i+2}+x_{i+3} \right),\\
\left( \dfrac{\partial x}{\partial \xi} \right)_i&=x_{i+1/2}-x_{i-1/2}.
\end{aligned}
\end{equation}
 The main difference between the two methods lies in how to handle the dissipation part  so that $\boldsymbol{D}^{+}$ and $\boldsymbol{D}^{-}$ vanish in uniform flows.

Nonomura et al. \cite{nonomura_new_2015} suggests that
all metrics in the dissipation terms are frozen
\begin{equation}\label{eq:nono_frozen_flux}
\begin{aligned}
\boldsymbol{\tilde{F}}_{m}&=\boldsymbol{F}_{m}\left(\dfrac{\xi_x}{J}\right)_{i+1/2}+\boldsymbol{G}_{m}\left(\dfrac{\xi_y}{J}\right)_{i+1/2}+\boldsymbol{H}_{m}\left(\dfrac{\xi_z}{J}\right)_{i+1/2},\\
\boldsymbol{\tilde{Q}}_{m}&=\boldsymbol{Q}_{m}\left(\dfrac{1}{J}\right)_{i+1/2},\quad m=i-2,\cdots,i+3,
\end{aligned}
\end{equation}
and
\begin{equation}\label{eq:nono_frozen}
\begin{aligned}
g_{i+1/2}=\dfrac{1}{2}\left( g_i+g_{i+1} \right),
\end{aligned}
\end{equation}
where the symbol $g$ denotes any of $\xi_x/J$, $\xi_y/J$, $\xi_z/J$ and $1/J$.
With respect to another method proposed by Zhu et al. \cite{zhu_free-stream_2019}, Eq. \eqref{eq:WENO_cen_dis_sub1} is rewritten into a local difference form, say $\hat{f}_{i,1}^{s,+} $,
\begin{equation} \label{eq:WENO_cen_dis_sub1_zhu1}
\begin{aligned}
\hat{f}_{i,1}^{s,+} =&\widetilde{f}_{i+1}^{s,+}-3\widetilde{f}_{i}^{s,+}+3\widetilde{f}_{i-1}^{s,+}-\widetilde{f}_{i-2}^{s,+} \\
=&\dfrac{1}{2} \boldsymbol{l}_{i+1/2}^s \left[  \left(\boldsymbol{\tilde{F}}_{i+1}-\boldsymbol{\tilde{F}}_{i}\right)-2\left(\boldsymbol{\tilde{F}}_{i}-\boldsymbol{\tilde{F}}_{i-1}\right)+\left(\boldsymbol{\tilde{F}}_{i-1}-\boldsymbol{\tilde{F}}_{i-2}\right) \right] \\
+& \dfrac{1}{2} \lambda^s \boldsymbol{l}_{i+1/2}^s \left[  \left(\boldsymbol{\tilde{Q}}_{i+1}-\boldsymbol{\tilde{Q}}_{i}\right)-2\left(\boldsymbol{\tilde{Q}}_{i}-\boldsymbol{\tilde{Q}}_{i-1}\right)+\left(\boldsymbol{\tilde{Q}}_{i-1}-\boldsymbol{\tilde{Q}}_{i-2}\right) \right].
\end{aligned}
\end{equation}
Then, the FP identity can be maintained by the following modifications,
\begin{equation} \label{eq:WENO_cen_dis_sub1_zhu2}
\begin{aligned}
\hat{f}_{i,1}^{s,+}
=&\dfrac{1}{2} \boldsymbol{l}_{i+1/2}^s \left[  \left(\boldsymbol{F}_{i+1}-\boldsymbol{F}_{i}\right)\left(\dfrac{\xi_x}{J}\right)_{i+1/2}-2\left(\boldsymbol{F}_{i}-\boldsymbol{F}_{i-1}\right)\left(\dfrac{\xi_x}{J}\right)_{i-1/2}+\left(\boldsymbol{F}_{i-1}-\boldsymbol{F}_{i-2}\right)\left(\dfrac{\xi_x}{J}\right)_{i-3/2} \right] \\
+&\dfrac{1}{2} \boldsymbol{l}_{i+1/2}^s \left[  \left(\boldsymbol{G}_{i+1}-\boldsymbol{G}_{i}\right)\left(\dfrac{\xi_y}{J}\right)_{i+1/2}-2\left(\boldsymbol{G}_{i}-\boldsymbol{G}_{i-1}\right)\left(\dfrac{\xi_y}{J}\right)_{i-1/2}+\left(\boldsymbol{G}_{i-1}-\boldsymbol{G}_{i-2}\right)\left(\dfrac{\xi_y}{J}\right)_{i-3/2} \right] \\
+&\dfrac{1}{2} \boldsymbol{l}_{i+1/2}^s \left[  \left(\boldsymbol{H}_{i+1}-\boldsymbol{H}_{i}\right)\left(\dfrac{\xi_z}{J}\right)_{i+1/2}-2\left(\boldsymbol{H}_{i}-\boldsymbol{H}_{i-1}\right)\left(\dfrac{\xi_z}{J}\right)_{i-1/2}+\left(\boldsymbol{H}_{i-1}-\boldsymbol{H}_{i-2}\right)\left(\dfrac{\xi_z}{J}\right)_{i-3/2} \right] \\
+& \dfrac{1}{2} \lambda^s \boldsymbol{l}_{i+1/2}^s \left[  \left(\boldsymbol{Q}_{i+1}-\boldsymbol{Q}_{i}\right)\left(\dfrac{1}{J}\right)_{i+1/2}-2\left(\boldsymbol{Q}_{i}-\boldsymbol{Q}_{i-1}\right)\left(\dfrac{1}{J}\right)_{i-1/2}+\left(\boldsymbol{Q}_{i-1}-\boldsymbol{Q}_{i-2}\right)\left(\dfrac{1}{J}\right)_{i-3/2} \right],
\end{aligned}
\end{equation}
where the cell-face metrics, say $g_{i+1/2}$, are evaluated by the 6th-order central scheme
 \begin{equation}\label{eq:half_metrics_6}
g_{i+1/2}=\dfrac{1}{60} \left( g_{i-2}-8g_{i-1}+37g_{i}+37g_{i+1}-8g_{i+2}+g_{i+3}  \right).
\end{equation}
Similarly, the smooth indicators can also be reformulated into a local difference form as well
\begin{equation}\label{eq:WENO_cen_dis_wet_zhu}
\begin{aligned}
\beta_0^{+}&= \dfrac{1}{4} \left[  \left(\widetilde{f}_{i-2}^{+}
                                 -\widetilde{f}_{i-1}^{+}\right)
                                 -3\left(\widetilde{f}_{i-1}^{+}-
                                   \widetilde{f}_{i}^{+} \right)\right]^2
            +\dfrac{13}{12}\left[ \left( \widetilde{f}_{i-2}^{+}
                                 -\widetilde{f}_{i-1}^{+}\right)
                               -\left(\widetilde{f}_{i-1}^{+}
                               -\widetilde{f}_{i}^{+} \right)\right]^2,
\end{aligned}
\end{equation}
where the same treatments as Eq. \eqref{eq:WENO_cen_dis_sub1_zhu2} can be applied.

Instead of operating in the central and dissipation part individually, Su et al. \cite{Su2020FP} give a straightforward strategy on FP by an approximation of the cell-center fluxes before the WENO reconstruction,
\begin{equation} \label{eq:my_point3}
\begin{aligned}
 \boldsymbol{\tilde{F}}_m \approx &\boldsymbol{\tilde{F}}_{m}-\boldsymbol{\tilde{F}}_{m}^*+\boldsymbol{\tilde{F}}_{i+1/2}^* \ , \quad m=i-2,\cdots,i+3 \\
 = & \left( \boldsymbol{F}_m-\boldsymbol{F^*} \right) \left(\dfrac{\xi_x}{J}\right)_m 
                                                                                     +\boldsymbol{F^*} \left(\dfrac{\xi_x}{J} \right)_{i+1/2} \\
														&+\left( \boldsymbol{G}_m-\boldsymbol{G^*} \right) \left(\dfrac{\xi_y}{J}\right)_m 
                                                                                     +\boldsymbol{G^*} \left(\dfrac{\xi_y}{J} \right)_{i+1/2} \\
														&+\left( \boldsymbol{H}_m-\boldsymbol{H^*} \right) \left(\dfrac{\xi_z}{J}\right)_m 
                                                                                     +\boldsymbol{H^*} \left(\dfrac{\xi_z}{J} \right)_{i+1/2},
\end{aligned}
\end{equation}
where $\boldsymbol{F^*}$, $\boldsymbol{G^*}$ and $\boldsymbol{H^*}$ are the reference state in the local stencil, as given in Ref. \cite{Su2020FP}.
Then, without modifying the original form of WENO scheme, the FP cell-face fluxes can be obtained directly by
\begin{equation} \label{eq:my_point2}
\begin{aligned}
 \boldsymbol{\tilde{F}}_{i+1/2}^{'}&=WENO \left( \boldsymbol{\tilde{F}}_{i-2}-\boldsymbol{\tilde{F}}_{i-2}^*+\boldsymbol{\tilde{F}}_{i+1/2}^*, \cdots, \boldsymbol{\tilde{F}}_{i+3}-\boldsymbol{\tilde{F}}_{i+3}^*+ \boldsymbol{\tilde{F}}_{i+1/2}^* \right).
\end{aligned}
\end{equation}

\section{The sufficient condition for FP in nonlinear upwind schemes}
In this section, we give a further study on the upwind dissipation and conclude a general rule for the previous FP method~\cite{nonomura_new_2015,zhu_free-stream_2019,Su2020FP}. After that, distinguished from the already existed sufficient condition of Deng et al. \cite{DENG201390} and Abe et al. \cite{abe2014geometric} for the linear schemes, a supplementary FP sufficient condition for the nonlinear upwind schemes is provided.
With this sufficient condition, the FP metrics are deduced to obtain a novel, simple and high-order strategy on FP for the WENO schemes. Without loss of generality, WENO5 is considered in the following elaborations.
\subsection{The general rule of the FP nonlinear upwind schemes}\label{sec:supp_cond}
The dissipation $\boldsymbol{ D }^+$ includes the combinations of cell-center conservative variables $\boldsymbol{ \tilde{Q} }_{m}$ and fluxes $\boldsymbol{ \tilde{F} }_{m}$, respectively, as shown in Eq. \eqref{eq:WENO_cen_dis_sub1}. Thus, we denote
\begin{equation}
\boldsymbol{D}^{+}=\boldsymbol{D_F}^{+}+\boldsymbol{D_Q}^{+},
\end{equation}
where
\begin{equation}\label{eq:DF_p_o}
\begin{aligned}
\boldsymbol{D_F}^{+}=&
-\dfrac{1}{60} \sum\limits_s \boldsymbol{r}_{i+1/2}^s  \left[ \dfrac{1}{2} \left(20\omega_0^+-1 \right)   \boldsymbol{l}_{i+1/2}^s \left(  \boldsymbol{\tilde{F}}_{i+1}-3\boldsymbol{\tilde{F}}_{i}+3\boldsymbol{\tilde{F}}_{i-1}-\boldsymbol{\tilde{F}}_{i-2}  \right)  \right.  \\
&\left. \qquad \qquad \qquad \quad -\dfrac{1}{2} \left( 10\omega_0^+ +10\omega_1^+ -5 \right)   \boldsymbol{l}_{i+1/2}^s \left(  \boldsymbol{\tilde{F}}_{i+2}-3\boldsymbol{\tilde{F}}_{i+1}+3\boldsymbol{\tilde{F}}_{i}-\boldsymbol{\tilde{F}}_{i-1} \right)  \right. \\
&\left. \qquad \qquad  \qquad \quad + \dfrac{1}{2} \boldsymbol{l}_{i+1/2}^s \left(  \boldsymbol{\tilde{F}}_{i+3}-3\boldsymbol{\tilde{F}}_{i+2}+3\boldsymbol{\tilde{F}}_{i+1}-\boldsymbol{\tilde{F}}_{i} \right)  \right],
\end{aligned}
\end{equation}
\begin{equation}\label{eq:DQ_p_o}
\begin{aligned}
\boldsymbol{D_Q}^{+}=&
-\dfrac{1}{60} \sum\limits_s \boldsymbol{r}_{i+1/2}^s \lambda^s \left[ \dfrac{1}{2} \left(20\omega_0^+-1 \right)   \boldsymbol{l}_{i+1/2}^s \left(  \boldsymbol{\tilde{Q}}_{i+1}-3\boldsymbol{\tilde{Q}}_{i}+3\boldsymbol{\tilde{Q}}_{i-1}-\boldsymbol{\tilde{Q}}_{i-2}  \right)  \right.  \\
&\left. \qquad \qquad \qquad \quad -\dfrac{1}{2} \left( 10\omega_0^+ +10\omega_1^+ -5 \right)   \boldsymbol{l}_{i+1/2}^s \left(  \boldsymbol{\tilde{Q}}_{i+2}-3\boldsymbol{\tilde{Q}}_{i+1}+3\boldsymbol{\tilde{Q}}_{i}-\boldsymbol{\tilde{Q}}_{i-1} \right)  \right. \\
&\left. \qquad \qquad  \qquad \quad + \dfrac{1}{2} \boldsymbol{l}_{i+1/2}^s \left(  \boldsymbol{\tilde{Q}}_{i+3}-3\boldsymbol{\tilde{Q}}_{i+2}+3\boldsymbol{\tilde{Q}}_{i+1}-\boldsymbol{\tilde{Q}}_{i} \right)  \right].
\end{aligned}
\end{equation}
It is observed that $\boldsymbol{D_Q}^{+}$ and $\boldsymbol{D_F}^{+}$ give the same form and it can be extended to the negative dissipation $\boldsymbol{D}^{-}$ (see Eq. \eqref{eq:WENO_cen_dis}) as well. Therefore, for the simplicity, we mainly focus on $\boldsymbol{D_Q}^{+}$ in the following discussions, which can be rewritten as
\begin{equation}\label{eq:DQ_p}
\begin{aligned}
\boldsymbol{D_Q}^{+}=&-\dfrac{1}{120} \sum\limits_s \boldsymbol{r}_{i+1/2}^s  \lambda^s \boldsymbol{l}_{i+1/2}^s \left[ \underbrace{-3 \boldsymbol{\tilde{P}}_{0} -9  \boldsymbol{\tilde{P}}_{1} + 9 \boldsymbol{\tilde{P}}_{2}+3 \boldsymbol{\tilde{P}}_{3}}_{\text{Linear dissipation}, \ \boldsymbol{D}_L}  \right.  \\
& \left. \qquad \qquad \qquad  + \underbrace{60\left( C_0-\omega_0^+ \right)   \boldsymbol{\tilde{P}}_{0} + 60\left( C_1-\omega_1^+ \right)\boldsymbol{\tilde{P}}_{1}  + 60\left( C_2-\omega_2^+ \right)\boldsymbol{\tilde{P}}_{2} }_{\text{nonlinear dissipation}, \ \boldsymbol{D}_N}  \right],
\end{aligned}
\end{equation}
where
\begin{equation}\label{eq:sub-stencil_P}
\begin{aligned}
\boldsymbol{\tilde{P}}_{0}&=\dfrac{1}{3}\boldsymbol{\tilde{Q}}_{i-2}-\dfrac{7}{6}\boldsymbol{\tilde{Q}}_{i-1}+\dfrac{11}{6}\boldsymbol{\tilde{Q}}_{i},\\
\boldsymbol{\tilde{P}}_{1}&=-\dfrac{1}{6}\boldsymbol{\tilde{Q}}_{i-1}+\dfrac{5}{6}\boldsymbol{\tilde{Q}}_{i}+\dfrac{1}{3}\boldsymbol{\tilde{Q}}_{i+1},\\
\boldsymbol{\tilde{P}}_{2}&=\dfrac{1}{3}\boldsymbol{\tilde{Q}}_{i}+\dfrac{5}{6}\boldsymbol{\tilde{Q}}_{i+1}-\dfrac{1}{6}\boldsymbol{\tilde{Q}}_{i+2},\\
\boldsymbol{\tilde{P}}_{3}&=\dfrac{11}{6}\boldsymbol{\tilde{Q}}_{i+1}-\dfrac{7}{6}\boldsymbol{\tilde{Q}}_{i+2}+\dfrac{1}{3}\boldsymbol{\tilde{Q}}_{i+3}.
\end{aligned}
\end{equation}
They indicate that the dissipation $\boldsymbol{D_Q}^{+}$ is composed by the linear and nonlinear parts, which are all the combinations of the four 3rd-order reconstructions.

Recalling to the FP method of Nonomura et al \cite{nonomura_new_2015}, they apply the frozen metrics in the dissipation, which results in
\begin{equation}\label{eq:sub-stencil_nono}
\begin{aligned}
\boldsymbol{\tilde{P}}_{0}^{N}&=\dfrac{1}{3}\boldsymbol{Q}_{i-2}g_{i+1/2}^{N}-\dfrac{7}{6}\boldsymbol{Q}_{i-1}g_{i+1/2}^{N}+\dfrac{11}{6}\boldsymbol{Q}_{i}g_{i+1/2}^{N},\\
\boldsymbol{\tilde{P}}_{1}^{N}&=-\dfrac{1}{6}\boldsymbol{Q}_{i-1} g_{i+1/2}^{N}+\dfrac{5}{6}\boldsymbol{Q}_{i}g_{i+1/2}^{N}+\dfrac{1}{3}\boldsymbol{Q}_{i+1} g_{i+1/2}^{N},\\
\boldsymbol{\tilde{P}}_{2}^{N}&=\dfrac{1}{3}\boldsymbol{Q}_{i}g_{i+1/2}^{N}+\dfrac{5}{6}\boldsymbol{Q}_{i+1}g_{i+1/2}^{N}-\dfrac{1}{6}\boldsymbol{Q}_{i+2}g_{i+1/2}^{N},\\
\boldsymbol{\tilde{P}}_{3}^{N}&=\dfrac{11}{6}\boldsymbol{Q}_{i+1}g_{i+1/2}^{N}-\dfrac{7}{6}\boldsymbol{Q}_{i+2}g_{i+1/2}^{N}+\dfrac{1}{3}\boldsymbol{Q}_{i+3} g_{i+1/2}^{N},
\end{aligned}
\end{equation}
where $g_{i+1/2}^{N}$ is calculated by Eq. \eqref{eq:nono_frozen}. 
Similarly, the dissipation in the FP method of Zhu et al \cite{zhu_free-stream_2019} can be reformulated as
\begin{equation}\label{eq:sub-stencil_zhu}
\begin{aligned}
\boldsymbol{\tilde{P}}_{0}^{Z}&=\dfrac{1}{3}\boldsymbol{Q}_{i-2}g_{i-3/2}^{Z}-\dfrac{7}{6}\boldsymbol{Q}_{i-1}\left(\dfrac{2}{7} g_{i-3/2}^{Z}+\dfrac{5}{7}g_{i-1/2}^{Z}\right)+\dfrac{11}{6}\boldsymbol{Q}_{i}\left(\dfrac{5}{11} g_{i-1/2}^{Z}+\dfrac{6}{11}g_{i+1/2}^{Z}\right),\\
\boldsymbol{\tilde{P}}_{1}^{Z}&=-\dfrac{1}{6}\boldsymbol{Q}_{i-1} g_{i-1/2}^{Z}+\dfrac{5}{6}\boldsymbol{Q}_{i}\left(\dfrac{1}{5} g_{i-1/2}^{Z}+\dfrac{4}{5}g_{i+1/2}^{Z}\right)+\dfrac{1}{3}\boldsymbol{Q}_{i+1} g_{i+1/2}^{Z},\\
\boldsymbol{\tilde{P}}_{2}^{Z}&=\dfrac{1}{3}\boldsymbol{Q}_{i}g_{i+1/2}^{Z}+\dfrac{5}{6}\boldsymbol{Q}_{i+1}\left(\dfrac{4}{5} g_{i+1/2}^{Z}+\dfrac{1}{5}g_{i+3/2}^{Z}\right)-\dfrac{1}{6}\boldsymbol{Q}_{i+2}g_{i+3/2}^{Z},\\
\boldsymbol{\tilde{P}}_{3}^{Z}&=\dfrac{11}{6}\boldsymbol{Q}_{i+1}\left(\dfrac{6}{11} g_{i+1/2}^{Z}+\dfrac{5}{11}g_{i+3/2}^{Z}\right)-\dfrac{7}{6}\boldsymbol{Q}_{i+2}\left(\dfrac{5}{7} g_{i+3/2}^{Z}+\dfrac{2}{7}g_{i+5/2}^{Z}\right)+\dfrac{1}{3}\boldsymbol{Q}_{i+3} g_{i+5/2}^{Z}.
\end{aligned}
\end{equation}
where $g_{i+1/2}^{Z}$ is calculated by Eq. \eqref{eq:half_metrics_6}. Besides, in the FP method of Su et al \cite{Su2020FP}, the dissipation can be given as
\begin{equation}\label{eq:sub-stencil_su}
\begin{aligned}
\boldsymbol{\tilde{P}}_{0}^{S}&=\dfrac{1}{3}\left(\boldsymbol{Q}_{i-2}-\boldsymbol{Q}^{*}\right)g_{i-2}-\dfrac{7}{6}\left(\boldsymbol{Q}_{i-1}-\boldsymbol{Q}^{*}\right)g_{i-1}+\dfrac{11}{6}\left(\boldsymbol{Q}_{i}-\boldsymbol{Q}^{*}\right)g_{i}+\boldsymbol{Q}^{*}g_{i+1/2}^{S},\\
\boldsymbol{\tilde{P}}_{1}^{S}&=-\dfrac{1}{6}\left(\boldsymbol{Q}_{i-1}-\boldsymbol{Q}^{*}\right) g_{i-1}+\dfrac{5}{6}\left(\boldsymbol{Q}_{i}-\boldsymbol{Q}^{*}\right)g_{i}+\dfrac{1}{3}\left(\boldsymbol{Q}_{i+1}-\boldsymbol{Q}^{*} \right)g_{i+1}+\boldsymbol{Q}^{*}g_{i+1/2}^{S},\\
\boldsymbol{\tilde{P}}_{2}^{S}&=\dfrac{1}{3}\left(\boldsymbol{Q}_{i}-\boldsymbol{Q}^{*}\right)g_{i}+\dfrac{5}{6}\left(\boldsymbol{Q}_{i+1}-\boldsymbol{Q}^{*}\right)g_{i+1}-\dfrac{1}{6}\left(\boldsymbol{Q}_{i+2}-\boldsymbol{Q}^{*}\right)g_{i+2}+\boldsymbol{Q}^{*}g_{i+1/2}^{S},\\
\boldsymbol{\tilde{P}}_{3}^{S}&=\dfrac{11}{6}\left(\boldsymbol{Q}_{i+1}-\boldsymbol{Q}^{*}\right)g_{i+1}-\dfrac{7}{6}\left(\boldsymbol{Q}_{i+2}-\boldsymbol{Q}^{*}\right)g_{i+2}+\dfrac{1}{3}\left(\boldsymbol{Q}_{i+3}-\boldsymbol{Q}^{*}\right) g_{i+3}+\boldsymbol{Q}^{*}g_{i+1/2}^{S},
\end{aligned}
\end{equation}
where $g_{i+1/2}^{S}$ can be calculated by SCMM with various central schemes, see Ref. \cite{Su2020FP}. 

According to Eqs. \eqref{eq:sub-stencil_nono}$\sim$\eqref{eq:sub-stencil_su}, a common rule of the above methods can be concluded by some simple algebraic operations, i.e.,
\begin{equation}\label{eq:sub-stencil_cond}
\begin{aligned}
\boldsymbol{\tilde{P}}_{0}=\boldsymbol{\tilde{P}}_{1}=\boldsymbol{\tilde{P}}_{2}=\boldsymbol{\tilde{P}}_{3}=\boldsymbol{Q}g_{i+1/2},
\end{aligned}
\end{equation}
under the uniform flow condition, where $\boldsymbol{Q}_m=\boldsymbol{Q}^*=\boldsymbol{Q}$ and $g_{i+1/2}$ stands for $g_{i+1/2}^N$, $g_{i+1/2}^Z$ or $g_{i+1/2}^S$.
Then, it is easy to prove from Eq. \eqref{eq:DQ_p} that this rule yields the linear and nonlinear dissipation terms reducing to zero, i.e.
\begin{equation}
\begin{aligned}
\boldsymbol{D}_L&=\boldsymbol{0},\\
\boldsymbol{D}_N&=\boldsymbol{0},
\end{aligned}
\end{equation}
which makes the free-stream condition being preserved for the upwind dissipation actually.
Moreover, we conduct a further investigation on the central part under the constraint of Eq. \eqref{eq:sub-stencil_cond}, i.e.,
\begin{equation}\label{sub_central}
\begin{aligned}
&\dfrac{1}{60} \left( \boldsymbol{\tilde{F}}_{i-2} -8 \boldsymbol{\tilde{F}}_{i-1}+37 \boldsymbol{\tilde{F}}_{i} +37\boldsymbol{\tilde{F}}_{i+1}-8 \boldsymbol{\tilde{F}}_{i+2}+\boldsymbol{\tilde{F}}_{i+3} \right)\\
=&\quad \dfrac{1}{20}\boldsymbol{\tilde{P}}_{0}+\dfrac{9}{20}\boldsymbol{\tilde{P}}_{1}+\dfrac{9}{20}\boldsymbol{\tilde{P}}_{2}+\dfrac{1}{20}\boldsymbol{\tilde{P}}_{3}\\
=&\quad \boldsymbol{F}\left(\dfrac{\xi_x}{J}\right)_{i+1/2}+\boldsymbol{G}\left(\dfrac{\xi_y}{J}\right)_{i+1/2}+\boldsymbol{H}\left(\dfrac{\xi_z}{J}\right)_{i+1/2},
\end{aligned}
\end{equation}
where $\boldsymbol{F}$, $\boldsymbol{G}$ and $\boldsymbol{H}$ are the constant fluxes of uniform flows.
Thus, the central part can preserve the free-stream condition as well, as long as $\left(\xi_x/J\right)_{i+1/2}, \cdots, \left(\xi_z/J\right)_{i+1/2}$ are evaluated by SCMM with the central scheme required by the sufficient condition of Deng et al. \cite{DENG201390} and Abe et al. \cite{abe2014geometric}.

According to this, we propose a supplementary sufficient condition for the nonlinear upwind schemes to achieve the FP identity as
\begin{remark}\label{remark_1}
if the metrics $g^*_{m},(m=i-2,\cdots,i+3)$ yields a unique value for the sub-stencils reconstructions, i.e.,
\begin{equation}\label{eq:SCL_condition}
\left \{
\begin{aligned}
   \dfrac{1}{3}g^*_{i-2}-\dfrac{7}{6}g^*_{i-1}+\dfrac{11}{6}g^*_{i}  &=g^*_{i+1/2}, \\
  -\dfrac{1}{6}g^*_{i-1}+\dfrac{5}{6}g^*_{i}+\dfrac{1}{3}g^*_{i+1} &=g^*_{i+1/2} ,\\
  \dfrac{1}{3}g^*_{i}+\dfrac{5}{6}g^*_{i+1}-\dfrac{1}{6}g^*_{i+2}   &=g^*_{i+1/2},\\
   \dfrac{11}{6}g^*_{i+1}-\dfrac{7}{6}g^*_{i+2}+\dfrac{1}{3}g^*_{i+3}  &=g^*_{i+1/2}, \\
\end{aligned}
\right.
\end{equation}
the FP identity can be maintained for the upwind dissipation of WENO5.
Moreover, if $g^*_{1+1/2}$
is evaluated by the SCMM with the 6th-order central scheme, the FP identity can be satisfied for the central part.
\end{remark}
This general condition gives a constraint for the metrics in the upwind schemes to remove the excrescent dissipation which results in the violation of the FP identity.
It is meaningful to guide the design of various FP upwind schemes.

\subsection{The novel FP method for WENO5 scheme}
A straightforward application of the above sufficient condition in Sec. \ref{sec:supp_cond}
is to develop new FP schemes.
Although the previous FP methods of Zhu et al. \cite{zhu_free-stream_2019} can obtain more accurate results than that in Ref. \cite{nonomura_new_2015} and \cite{Su2020FP}, it yields more complexities in implementation due to the inconformity of the modified metrics in each sub-stencil, as shown in Eq.~\eqref{eq:sub-stencil_zhu}, which requires to be achieved in a local difference form with the original form of the upwind schemes being changed.
Thus, we propose the new FP metrics denoted by $g_{i+1/2}^{*},g_{i-2}^*,\cdots,g_{i+3}^{*}$ 
\begin{equation}\label{eq:FP_metrics}
\left \{
\begin{aligned}
g^*_{i}&=g_{i}, \quad g^*_{i+1/2}=g_{i+1/2}, \quad g^*_{i+1}=g_{i+1},  \\
\dfrac{1}{6}g^*_{i-1}&=\dfrac{5}{6}g^*_{i}+\dfrac{1}{3}g^*_{i+1}-g^*_{i+1/2}, \\
\dfrac{1}{3}g^*_{i-2}&=\dfrac{7}{6}g^*_{i-1}-\dfrac{11}{6}g^*_{i}+g^*_{i+1/2}, \\
\dfrac{1}{6}g^*_{i+2}&=\dfrac{5}{6}g^*_{i+1}+\dfrac{1}{3}g^*_{i}-g^*_{i+1/2}, \\
\dfrac{1}{3}g^*_{i+3}&=\dfrac{7}{6}g^*_{i+2}-\dfrac{11}{6}g^*_{i+1}+g^*_{i+1/2} .
\end{aligned}
 \right.
\end{equation}
where $g_{m}$ is computed by the SCMM in Eq.~\eqref{eq:metric_discret} and $g_{i+1/2}$ is calculated by the 6th-order central scheme in Eq.~\eqref{eq:half_metrics_6}. Actually, Eq. \eqref{eq:SCL_condition} can be satisfied under these definitions, making the FP identity being satisfied for the WENO5 schemes. Unlike the previous method, the present FP method adopts the FP metrics to reconstruct the cell-face fluxes $\boldsymbol{\tilde{F}}^*_{i+1/2}$ directly without modify the original form of WENO scheme.

However, $\boldsymbol{\tilde{F}}^*_{i+1/2}$ computed with  $g_m^*$ only achieves a 3rd-order accuracy since $g^*_m$ ($m=i-2, i-1,i+2,i+3$) are reconstructed by the 3rd-order scheme from the original 6th-order central ones $g_m$ ($m=i, i+1/2, i+1$) (see Appendix A).
In the split form of the original nonlinear upwind schemes,
%
\begin{equation}\label{eq:Flux_c6d5}
\begin{aligned}
  \boldsymbol{\tilde{F}}_{i+1/2} &= \boldsymbol{\tilde{F}}^{C6}_{i+1/2} +
  \boldsymbol{\tilde{D}}^{U5}_{i+1/2},\\
\boldsymbol{\tilde{F}}^{C6}_{i+1/2} &= \dfrac{1}{60}\left( \boldsymbol{\tilde{F}}_{i-2}-8\boldsymbol{\tilde{F}}_{i-1}+37\boldsymbol{\tilde{F}}_{i}+37\boldsymbol{\tilde{F}}_{i+1}-8\boldsymbol{\tilde{F}}_{i+2}+\boldsymbol{\tilde{F}}_{i+3} \right),
\end{aligned}
\end{equation}
the central part $\tilde{F}^{C6}_{i+1/2}$ and the upwind dissipation $\tilde{D}^{U5}_{i+1/2}$ maintain 6th-order and 5th-order accuracy, respectively, since $g_m$ is applied.

Similarly, for the above FP upwind schemes, we have
\begin{equation}\label{eq:Flux_c3d5}
\begin{aligned}
\boldsymbol{\tilde{F}}^{*}_{i+1/2} &= \boldsymbol{\tilde{F}}^{*,C3}_{i+1/2} +
\boldsymbol{\tilde{D}}^{*}_{i+1/2},\\
\boldsymbol{\tilde{F}}^{*,C3}_{i+1/2} &= \dfrac{1}{60}\left(
        \boldsymbol{\tilde{F}}^{*}_{i-2}
     -8 \boldsymbol{\tilde{F}}^{*}_{i-1}
     +37\boldsymbol{\tilde{F}}^{*}_{i}
     +37\boldsymbol{\tilde{F}}^{*}_{i+1}
     -8 \boldsymbol{\tilde{F}}^{*}_{i+2}
       +\boldsymbol{\tilde{F}}^{*}_{i+3} \right).
\end{aligned}
\end{equation}
Although $\boldsymbol{\tilde{F}}^{*,C3}_{i+1/2}$ which is the central part of $\boldsymbol{\tilde{F}}^*_{i+1/2}$ reduces to the 3rd-order accuracy, we can prove (in Appendix A) that
\begin{remark}\label{remark_2}
the 3rd-order accurate FP metrics given in Eq.~\eqref{eq:FP_metrics} do not change the 5th-order accuracy of the upwind dissipation $\boldsymbol{\tilde{D}}^*_{i+1/2}$ reconstructed by the LU5 or WENO5 scheme.
\end{remark}
Thus, we suggest compensating  the accuracy loss of $\boldsymbol{\tilde{F}}^{*,C3}_{i+1/2}$ by
replacing it with ${\tilde{F}}^{C6}_{i+1/2}$
\begin{equation}\label{eq:Flux_cor}
\boldsymbol{\tilde{F}}^{'}_{i+1/2}= \boldsymbol{\tilde{F}}^*_{i+1/2}+\left(\boldsymbol{\tilde{F}}^{C6}_{i+1/2}-\boldsymbol{\tilde{F}}^{*,C3}_{i+1/2}\right).
\end{equation}
Thus, the new fluxes $\boldsymbol{\tilde{F}}_{i+1/2}^{'}$ can approach the 5th-order accuracy at non-critical points.

As a conclusion, the new FP WENO5 method is given by the following steps:
\begin{enumerate}
\item Obtain the FP metrics and Jacobian $g_m^*$ by Eq.~\eqref{eq:FP_metrics} in which $g_m$ should have been calculated by Eq.~\eqref{eq:metric_discret} beforehand;
\item Apply $g_m^*$ and $g_m$ to compute the cell center fluxes $\boldsymbol{\tilde{F}}_{m}^*$ and $\boldsymbol{\tilde{F}}_{m}$, respectively;
\item Employ the specific upwind schemes to reconstruct the cell-face fluxes $\boldsymbol{\tilde{F}}_{i+1/2}^*$, and calculate $\boldsymbol{\tilde{F}}^{C6}_{i+1/2}$ and $\boldsymbol{\tilde{F}}^{*,C3}_{i+1/2}$ by Eqs. \eqref{eq:Flux_c6d5} and \eqref{eq:Flux_c3d5};
\item Obtain the final cell-face fluxes $\boldsymbol{\tilde{F}}_{i+1/2}^{'}$ by Eq.~\eqref{eq:Flux_cor}.
\end{enumerate}
It should be pointed out that the added computational cost overhead here is small since only a few extra linear combinations, such as $g^*$, $\boldsymbol{\tilde{F}}^{C6}_{i+1/2}$ and $\boldsymbol{\tilde{F}}^{*,C3}_{i+1/2}$, have to be computed for each face in this method. In addition, with the sufficient condition, it is straightforward to extend this method to the 7th-order WENO scheme (see Appendix B).

\section{Numerical tests on curvilinear grids}
Various verifications, such as the isotropic vortex convection, the double Mach reflection problem, the transonic flow past the ONERA M6 wing, etc. are conducted to evaluate the accuracy and the capability in shock capturing of the proposed FP method on curvilinear grids. If not otherwise specified, the local Lax-Friedrichs flux splitting and the 3rd-order TVD Runge-Kutta scheme~\cite{gottlieb1998total} are utilized for the simulations. For the viscous issues, the 6th-order central scheme is adopted to discrete the viscous terms. In the followings, WENO5, WENO7 stand for the original 5th- and 7th-order WENO schemes of Shu~\cite{shu1998essentially}, WENOZ is the original improved 5th-order WENO scheme of Borges et al.~\cite{borges_improved_2008} and WENO5-Present, WENO7-Present, WENOZ-Present are the FP schemes suggested in the present paper.

\subsection{\label{sec:simple}Free-stream preserving and accuracy verifications}
\subsubsection{\label{sec:Free-stream}Free-stream}
\begin{figure} [htbp]
    \center
    \includegraphics[width=0.8\textwidth]{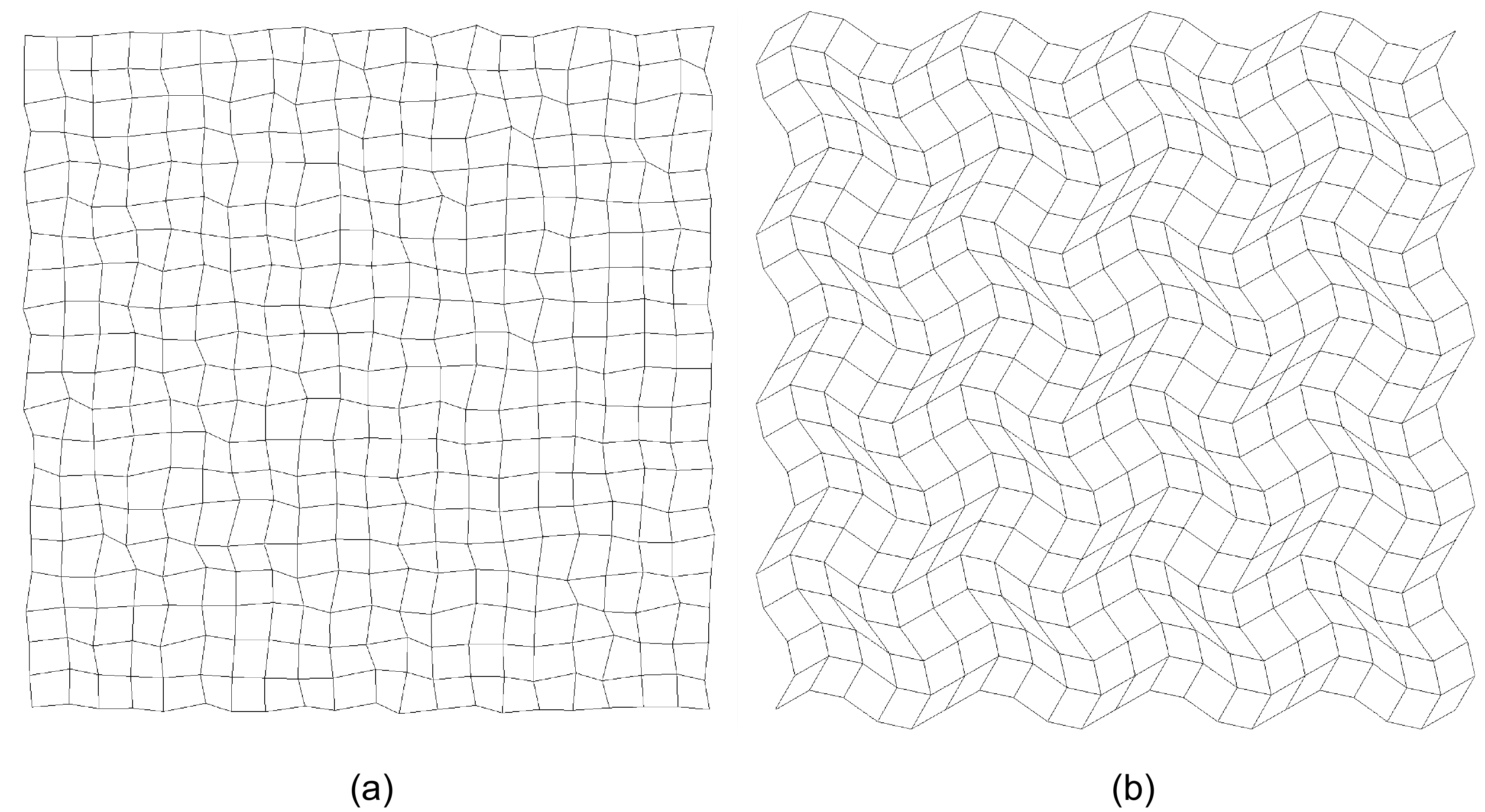}
    \caption{Illustrate the randomized (a) and wavy (b) grids. }
    \label{grid41}
\end{figure}

In the first case, the FP identity of the present schemes is tested on the wavy and randomized grids with a resolution of $21\times21$, as shown in Fig.~\ref{grid41}, respectively. The wavy grid is defined in the domain $(x,y)\in [-10,10]\times[-10,10]$ by
\begin{equation}\label{eq:wavy_grid}
\begin{aligned}
x_{i,j}=x_{min}+\Delta{x_0} \left[ (i-1)+A_x sin \left( \dfrac{n_{xy}\pi(j-1)\Delta{y_0}}{L_y} \right)\right] \\
y_{i,j}=y_{min}+\Delta{y_0} \left[ (j-1)+A_y sin \left( \dfrac{n_{yx}\pi(i-1)\Delta{x_0}}{L_x} \right)\right]
\end{aligned}
\end{equation}
where $L_x=L_y=20$, $x_{min}=-L_x/2$, $y_{min}=-L_y/2$, $A_x\Delta{x}=0.6$, $A_y\Delta{y}=0.6$, and $n_{xy}=n_{yx}=8$. 
And the randomized grid is disturbed randomly with $20\%$ grid spacing of the uniform Cartesian grid.

The uniform flows of $M=0.5$ in x direction is given as
\begin{equation}
u=0.5,v=0,p=1,\rho=\gamma
\end{equation}
where $\gamma=1.4$ is the specific heat ratio. The average ($L_2$-norm) and maximum ($L_\infty$-norm) errors of the velocity component $v$ for the two grids are estimated at $t=20$. As listed in Table~\ref{freestream_L2}, in contrast to the original WENO5, WENOZ and WENO7 schemes which bring large FP violated errors under the uniform flow, the proposed FP method effectively decreases the geometrically induced errors below $1\times10^{-14}$, being close to the machine zero for the double-precision computations.

\begin{table}[htbp]
   \centering
	\caption{The $L_2$ and $L_{\infty}$ errors of the $v$ component on the wavy and randomized grids.}

	\begin{tabular}{ccccc}
   \hline
    \multirow{2}*{Method}  & \multicolumn{2}{c}{Wavy grid}   &\multicolumn{2}{c}{Random grid} \\
   \cline{2-5}  
    ~ &$L_2$ error  &$L_{\infty}$ error &$L_2$ error  &$L_{\infty}$ error  \\
   \hline  
   WENO5         &$2.45\times10^{-2}  $  &$4.72\times10^{-2}$  &$1.29\times10^{-2}$    &$4.41\times10^{-2}$ \\
   WENOZ         &$6.53\times10^{-3}  $  &$1.32\times10^{-2}$  &$4.97\times10^{-3}$    &$1.65\times10^{-2}$ \\
   WENO7         &$1.03\times10^{-2}  $  &$1.98\times10^{-2}$  &$1.57\times10^{-2}$    &$5.08\times10^{-2}$ \\
   WENO5-Present   &$6.00\times10^{-16}$ &$2.13\times10^{-15}$ &$7.88\times10^{-16}$  &$2.12\times10^{-15}$ \\
   WENOZ-Present   &$1.10\times10^{-15}$ &$3.17\times10^{-15}$ &$1.75\times10^{-15}$  &$5.01\times10^{-15}$ \\
   WENO7-Present   &$5.49\times10^{-16}$ &$1.74\times10^{-15}$ &$6.19\times10^{-16}$  &$1.95\times10^{-15}$ \\
		\hline  
	\end{tabular}
	\label{freestream_L2}
\end{table}

\subsubsection{\label{sec:Isotropic vortex}Isotropic vortex}
To evaluate the accuracy and vortex preserving capability of the present FP schemes, the two-dimensional(2D) moving isotropic vortex problems on the wavy and randomized grids are investigated. This problem superposes an isotropic vortex centered at $(x_c,y_c)=(0,0)$ to a uniform flow with a Mach number of 0.5. Specifically, the perturbations of the velocity, temperature and entropy are expressed by:
\begin{equation}
\begin{aligned}
&(\delta u,\delta v)=\epsilon \tau e^{\alpha(1-\tau^2)} (sin{\theta},-cos{\theta}) \\
&\delta{T}=-\dfrac{(\gamma-1)\epsilon^2}{4\alpha\gamma}e^{2\alpha(1-\tau^2)} \\
&\delta{S}=0
\end{aligned}
\end{equation}
where $\alpha=0.204$, $\tau=r/r_c$ and $r=\sqrt{(x-x_c)^2+(y-y_c)^2}$. 
$r_c=1.0$, $\epsilon=0.02$ denote the vortex core length and strength, respectively. 
$T=p/\rho$ is the temperature, $S=p/\rho^\gamma$ is the entropy and $\gamma=1.4$. 
The periodic boundary condition is imposed and the results are estimated when the vortex moving back to the original position at $t=40$.

\begin{figure} [htbp]
    \center
     \includegraphics[width=0.58\textwidth]{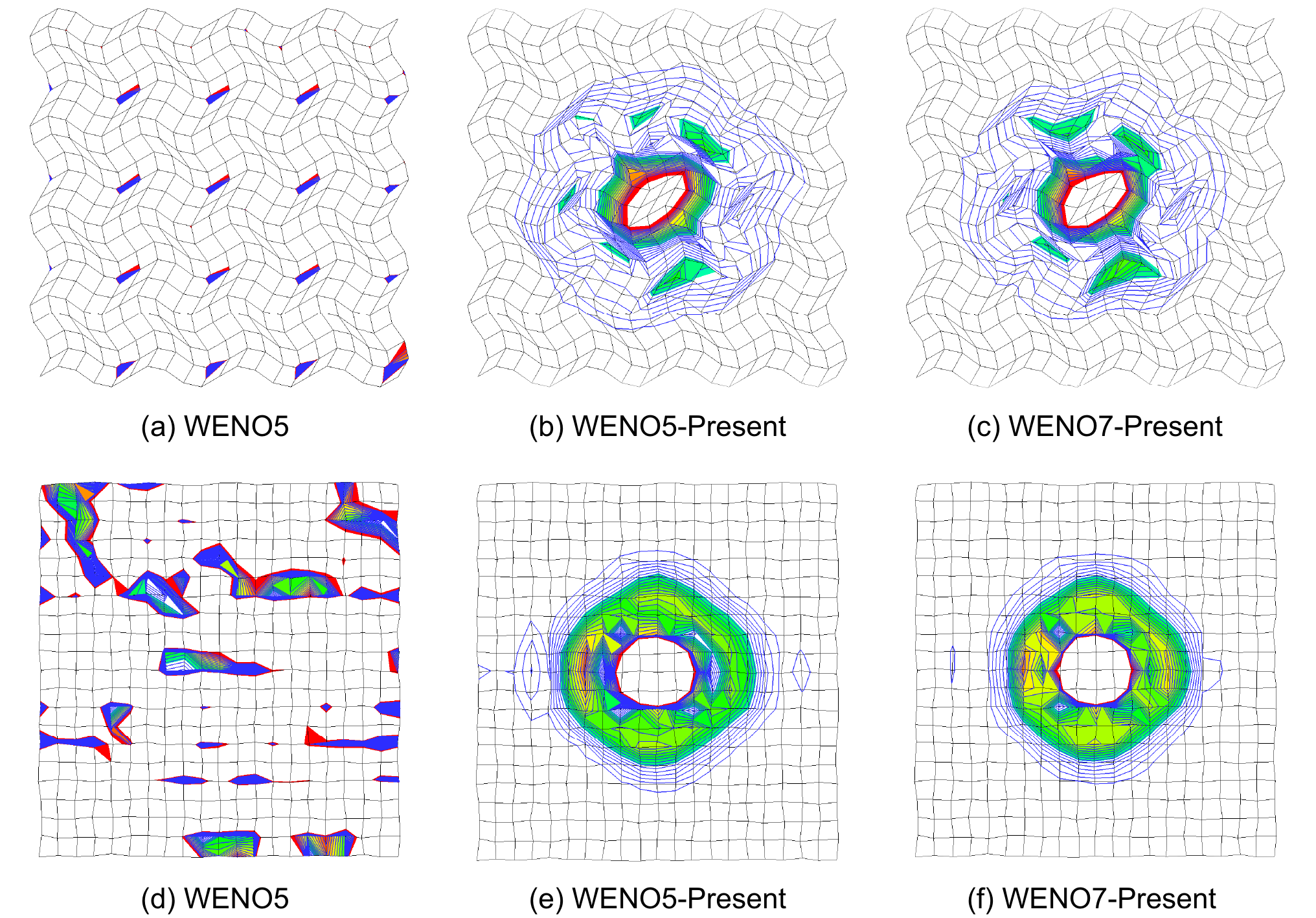}
    \caption{The vorticity magnitude distributions ranging from 0.0 to 0.006 of the 2D moving vortex on the wavy and randomized grids ($21\times 21$) for the WENOZ scheme.}
    \label{vortex_contours1}
        \includegraphics[width=0.8\textwidth]{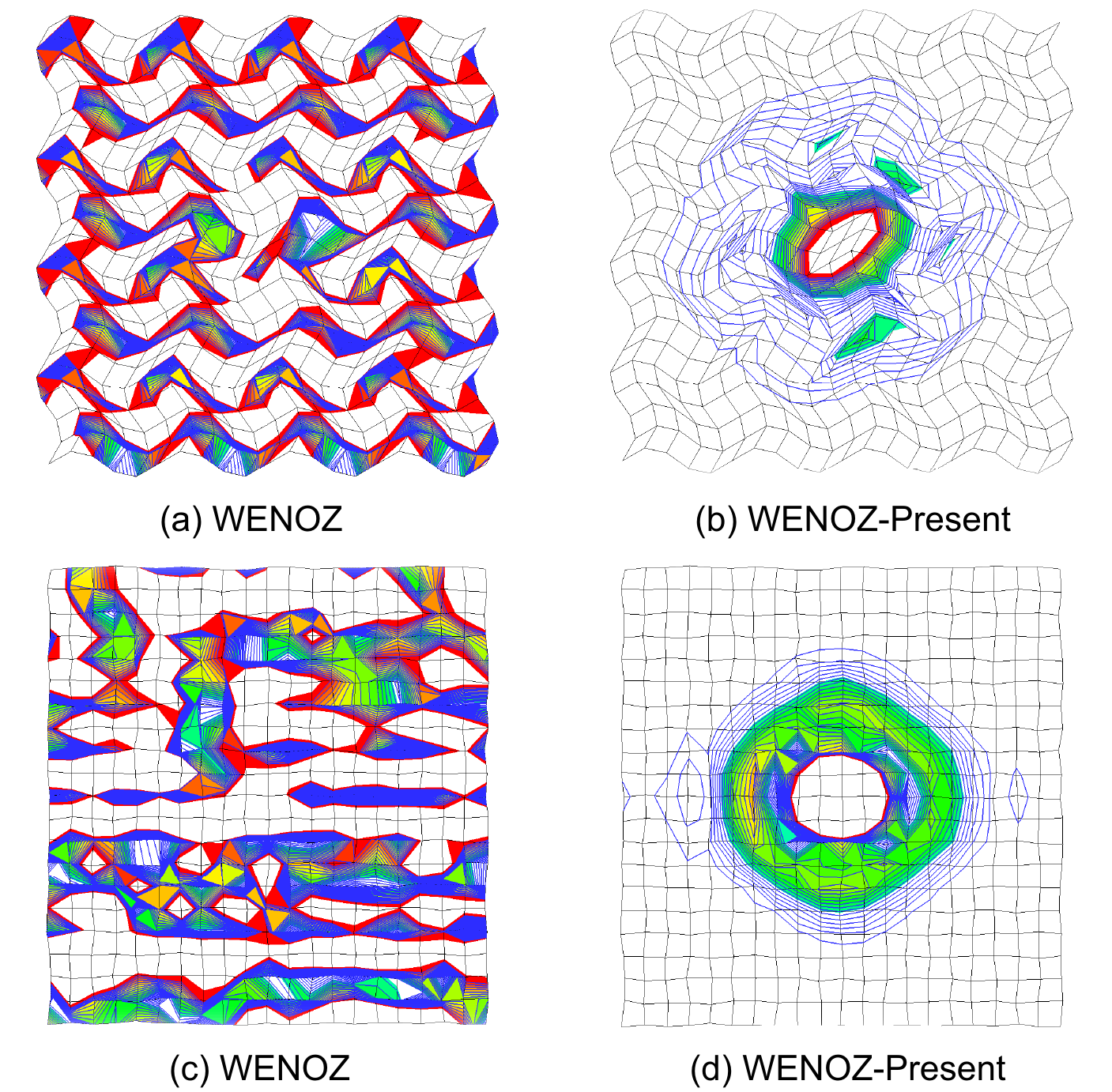}
    \caption{The vorticity magnitude distributions ranging from 0.0 to 0.006 of the 2D moving vortex on the wavy and randomized grids ($21\times 21$) for the WENO5 and WENO7 schemes.}
    \label{vortex_contours2}
\end{figure}

In Fig.~\ref{vortex_contours1} and Fig.~\ref{vortex_contours2}, the numerical vorticity magnitude contours on the wavy and random grids at a resolution of $21\times21$ indicate that the moving vortexes on those inhomogeneous grid can not be resolved for the original WENO5 and WENOZ schemes, while they can be well resolved for the present FP schemes (WENO5-Present, WENOZ-Present and WENO7-Present). With the goal of evaluating the grid convergence rate of these schemes, five grids with the resolution of $21\times21$, $41\times41$, $81\times81$, $161\times161$ and $321\times321$, respectively, are considered to test their accuracy. According to the previous studies~\cite{nonomura_new_2015,zhu_free-stream_2019}, the time step $\Delta{t}$ respect to each grid decreases until the $L_2$ and $L_{\infty}$ errors are invariant, aimed to eliminate the errors from the 3rd-order time integration. The $L_2$ and $L_{\infty}$ errors of the $v$ component and the corresponding convergence rates on the wavy grids, listed in Table~\ref{wavy_grid}, prove that the present FP schemes can achieve their theoretical convergence orders normally. Moreover, though the original WENO schemes provide small errors under a high resolution grid ($321\times321$), the present FP schemes show a better performance than them.
\begin{table}[htbp]
   \centering
	\caption{The $L_2$ and $L_{\infty}$ errors of the $v$ component and their corresponding convergence rates on the wavy grids. }
	\begin{tabular}{cccccc}
	    \hline
    Method &Grid size &$L_2$ error  &Convergence rate  &$L_{\infty}$ error  &Convergence rate \\
        \hline  
        WENO5 & $21 \times 21$ &$2.14\times10^{-2}$ &- &$5.06\times10^{-2}$  &- \\
		          &$41\times 41$ &$2.62\times10^{-3}$ &3.03 &$9.32\times10^{-3}$ &2.44 \\
		          &$81\times 81$ &$1.71\times10^{-4}$ &3.94 &$5.61\times10^{-4}$  &4.05 \\
		          &$161\times 161$ &$3.00\times10^{-6}$ &5.83 &$1.89\times10^{-5}$  &4.89 \\
		          &$321\times 321$ &$4.33\times10^{-8}$ &6.11 &$4.53\times10^{-7}$ &5.38 \\
       \hline  
    WENOZ       &$21 \times 21$    &$7.90\times10^{-3}$ &-       &$2.43\times10^{-2}$  &- \\
		        &$41\times 41$     &$8.34\times10^{-4}$ &3.24    &$4.98\times10^{-3}$  &2.29 \\
		        &$81\times 81$     &$3.92\times10^{-5}$ &4.41    &$1.94\times10^{-4}$  &4.68 \\
		        &$161\times 161$   &$1.23\times10^{-6}$ &4.99    &$1.05\times10^{-5}$  &4.21 \\
	  	        &$321\times 321$   &$3.89\times10^{-8}$ &4.98    &$4.10\times10^{-7}$  &4.68 \\
       \hline  
	WENO5-Present&$21 \times 21$    &$2.29\times10^{-3}$ &-       &$1.61\times10^{-2}$  &- \\
		        &$41\times 41$     &$4.82\times10^{-4}$ &2.25    &$4.36\times10^{-3}$  &1.88 \\
		        &$81\times 81$     &$1.66\times10^{-5}$ &4.86    &$1.47\times10^{-4}$  &4.89 \\
		        &$161\times 161$   &$5.85\times10^{-7}$ &4.83    &$5.82\times10^{-6}$  &4.66 \\
	  	        &$321\times 321$   &$1.96\times10^{-8}$ &4.90    &$2.17\times10^{-7}$  &4.75 \\
       \hline  
   WENOZ-Present&$21 \times 21$    &$2.31\times10^{-3}$ &-       &$1.58\times10^{-2}$  &- \\
		        &$41\times 41$     &$5.23\times10^{-4}$ &2.14    &$4.45\times10^{-3}$  &1.83 \\
		        &$81\times 81$     &$1.91\times10^{-5}$ &4.78    &$1.99\times10^{-4}$  &4.48 \\
		        &$161\times 161$   &$5.89\times10^{-7}$ &5.02    &$5.76\times10^{-6}$  &5.11 \\
	  	        &$321\times 321$   &$1.96\times10^{-8}$ &4.91    &$2.17\times10^{-7}$  &4.73 \\
       \hline  
   WENO7-Present&$21 \times 21$    &$2.16\times10^{-3}$ &-       &$1.49\times10^{-2}$  &- \\
		        &$41\times 41$     &$4.37\times10^{-4}$ &2.31    &$3.88\times10^{-3}$  &1.94 \\
		        &$81\times 81$     &$3.71\times10^{-6}$ &6.88    &$3.83\times10^{-5}$  &6.66 \\
		        &$161\times 161$   &$3.71\times10^{-8}$ &6.64    &$4.65\times10^{-7}$  &6.36 \\
                &$321\times 321$   &$3.12\times10^{-10}$&6.89    &$4.44\times10^{-9}$  &6.71 \\
		\hline  
	\end{tabular}
	\label{wavy_grid}
\end{table}
\subsection{\label{sec:invicid}Inviscid supersonic shock-capturing problems}
\subsubsection{\label{sec:DM}Double Mach reflection}
\begin{figure} 
    \center
    \includegraphics[width=0.82\textwidth]{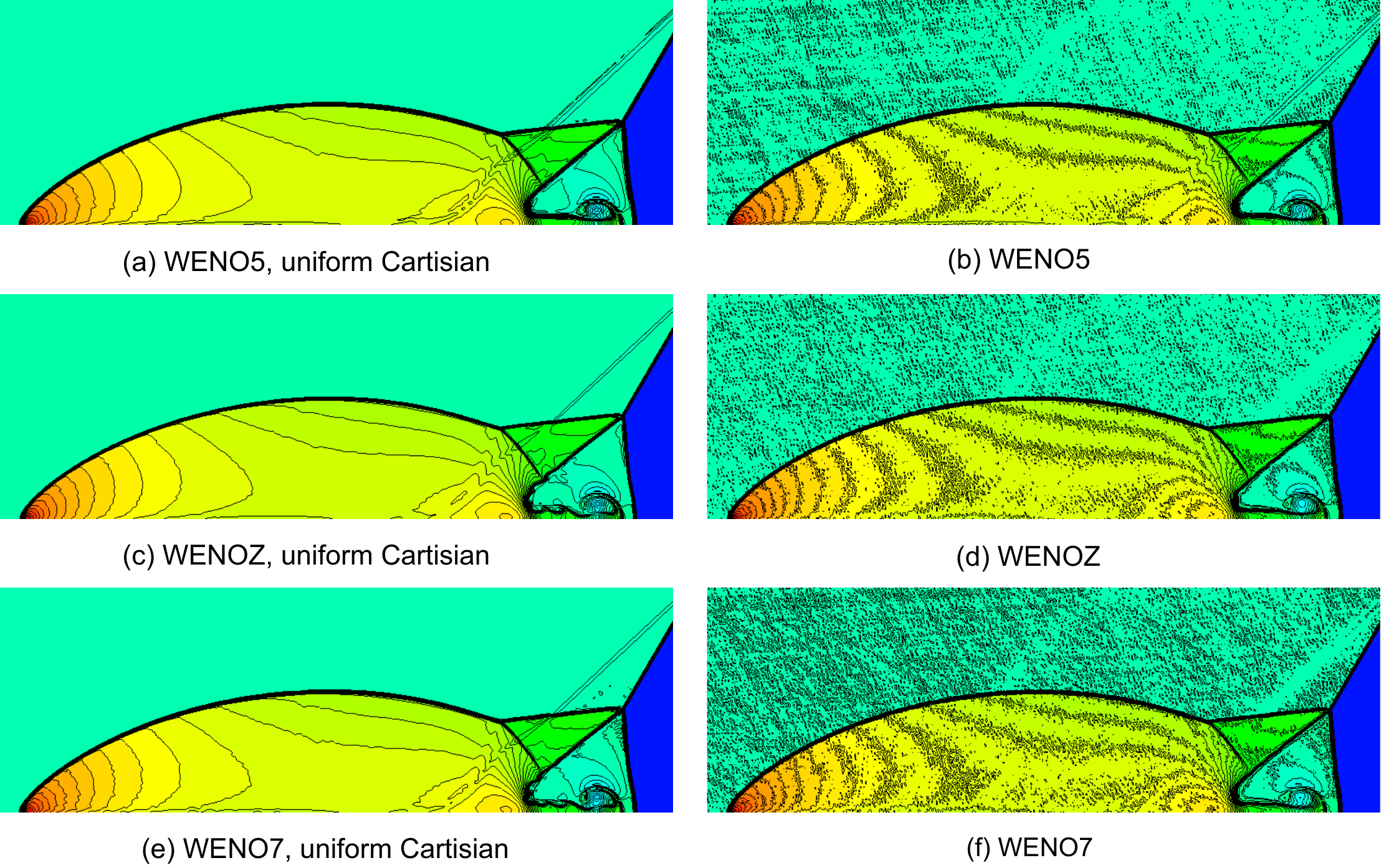}
    \caption{The density contours of the double Mach reflection problem ranging from 1.25 to 21.5 of the standard WENO schemes on the uniform Cartisian and $5\%$ randomized grid, respectively.}
    \label{double_mach1_1}
\end{figure}

\begin{figure} 
    \center
    \includegraphics[width=0.6\textwidth]{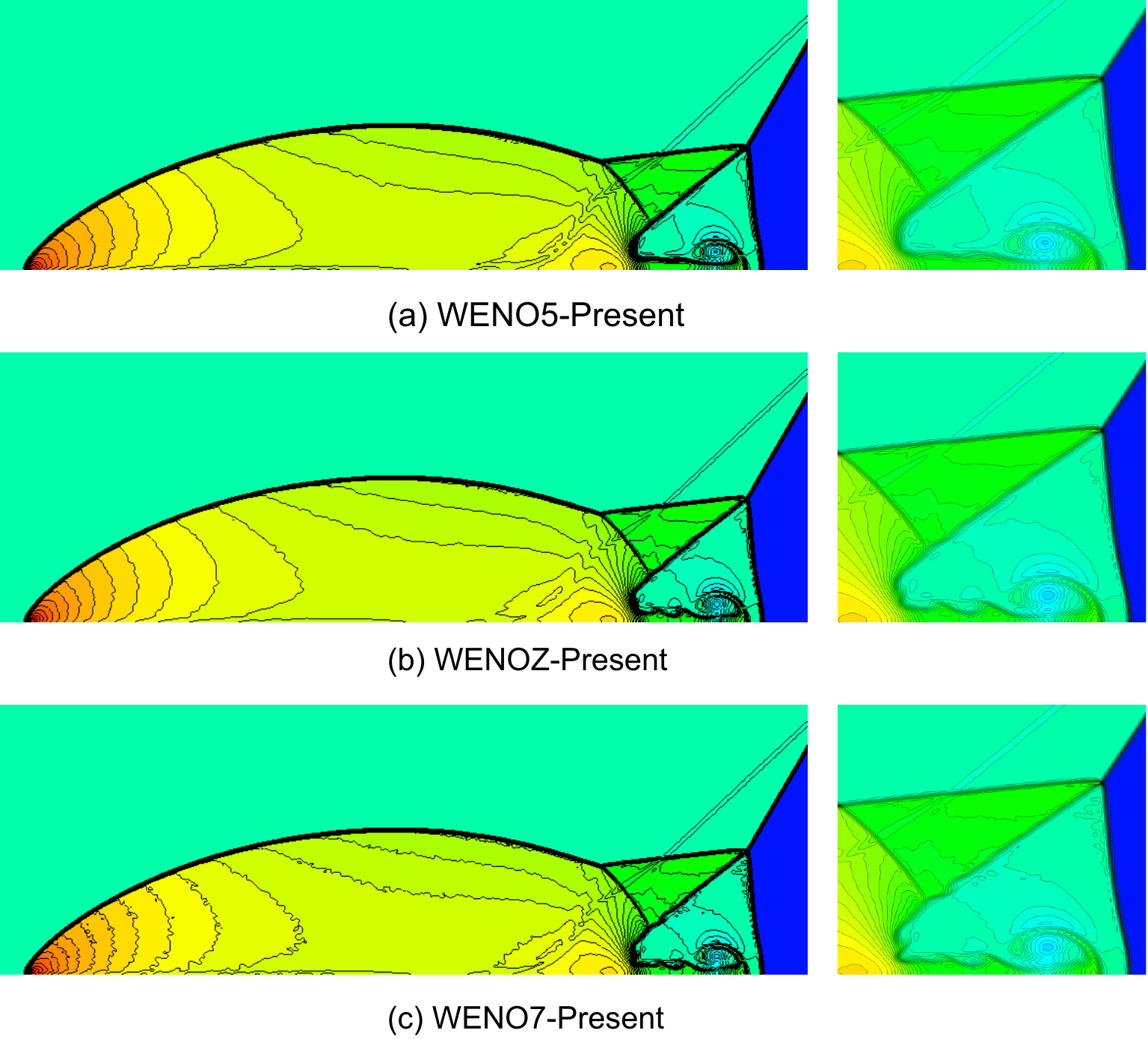}
    \caption{The density contours of the double Mach reflection problem ranging from 1.25 to 21.5 of the present WENO schemes on the $5\%$ randomized grid.}
    \label{double_mach1_2}
\end{figure}

\begin{figure} 
    \center
    \includegraphics[width=\textwidth]{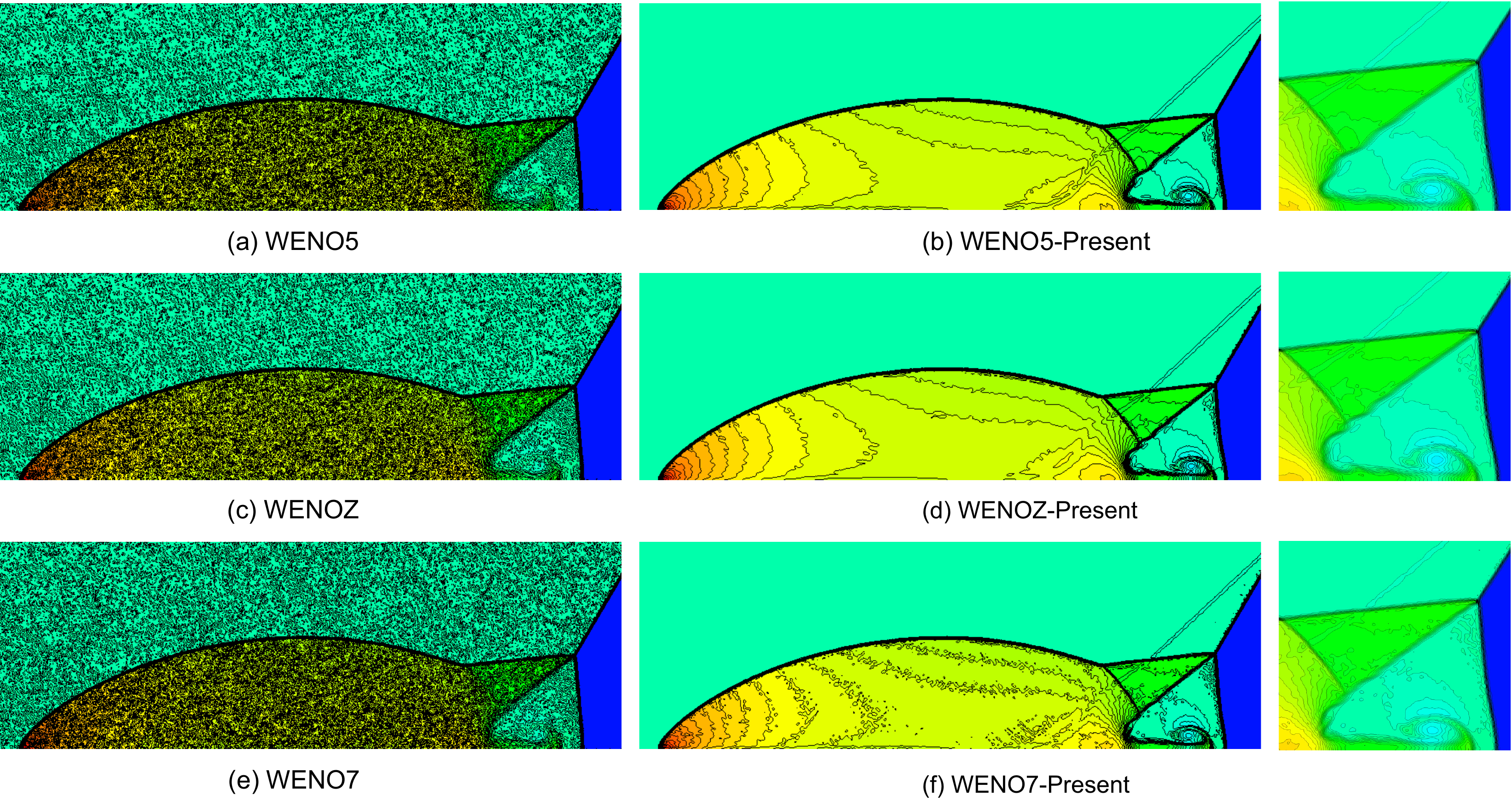}
    \caption{The density contours of the double Mach reflection problem ranging from 1.25 to 21.5 on the $20\%$ randomized grid.}
    \label{double_mach2}
\end{figure}

First, the double Mach problem~\cite{woodward_numerical_1984} is carried out to demonstrate the shock-capturing capability of the present free-stream preserving WENO scheme. The initial condition is
\begin{equation}
\begin{aligned}
 \left( \rho,u,v,p \right)=
 \begin{cases}
 (1.4,0,0,1.0)                & if \quad x-y tan(\pi/6) \geq 1/6 ,  \\
 (8.0, 7.1447, -4.125, 116.5) & if \quad x-y tan(\pi/6) < 1/6 .
 \end{cases}
\end{aligned}
\end{equation}
The computation is conducted up to $t=0.2$ under a CFL number of 0.5 and with the global Lax-Friedrichs dissipation. As illustrated in Fig.~\ref{double_mach1_1}, the calculated density contours of the double Mach reflection on the $5\%$ randomized grid with a resolution of $961\times241$ show that the original WENO5, WENOZ and WENO7 schemes induce spurious oscillations even for the smooth regions due to the violation of the FP identity. In contrast, WENO5-Present, WENOZ-Present and WENO7-Present remove the disturbances so that the shocks and continuous flows are resolved sharply and smoothly, respectively, see Fig.~\ref{double_mach1_2}. When the grid is randomized up to $20\%$ of the uniform grid spacing, as given in Fig.~\ref{double_mach2}, the density contours of the original WENO schemes are filled with noises completely. However, the results calculated by the proposed FP WENO schemes show very few non-physical oscillations. Though some disturbances are observed in the contour of WENO7-Present, it is essentially improved compared with the original ones.
\subsubsection{\label{sec:sphere}Supersonic flow past a cylinder}
\begin{figure} [htbp]
    \center
    \includegraphics[width=\textwidth]{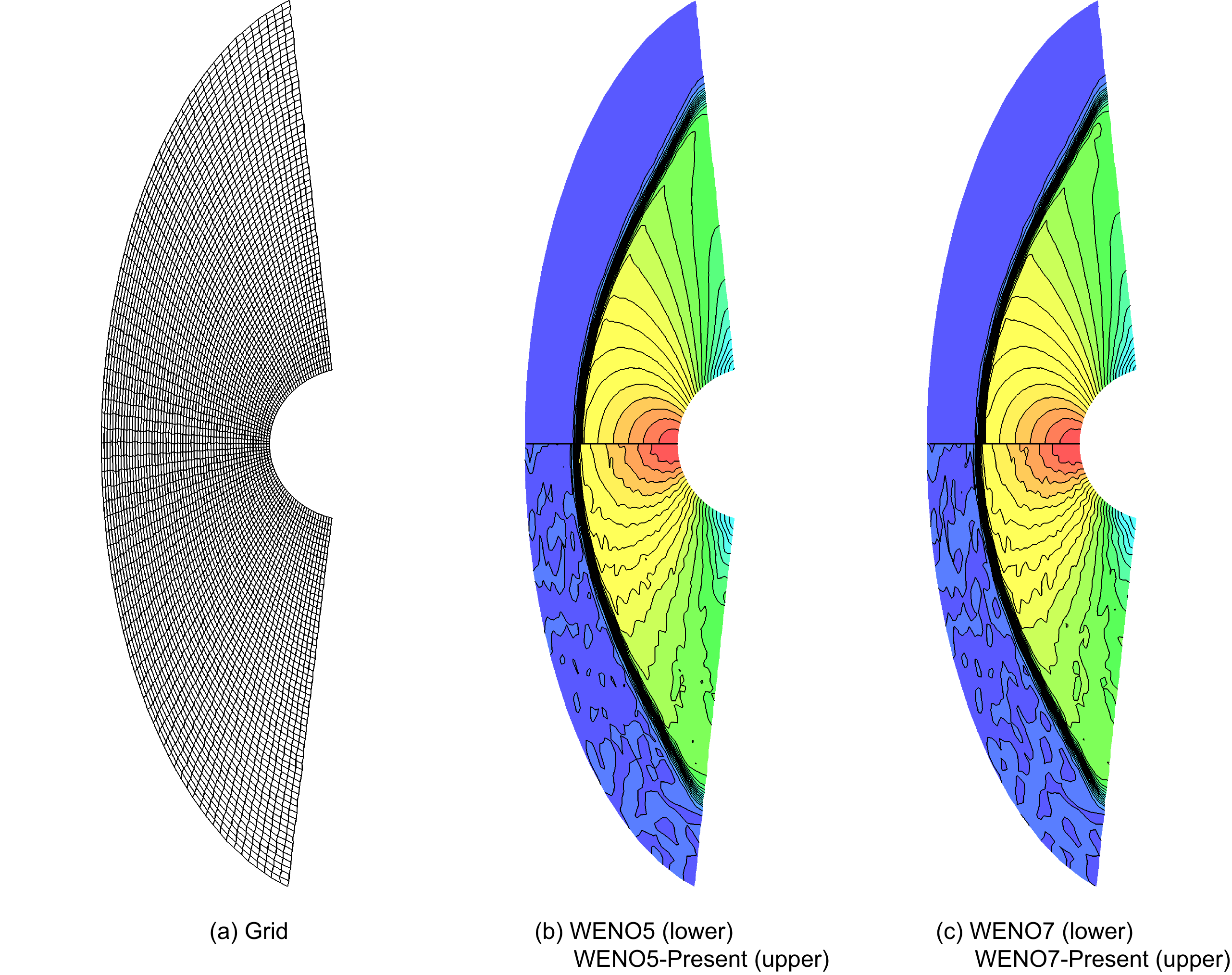}
    \caption{The pressure distributions from 1.0 to 5.5 of the supersonic flow past a cylinder. }
    \label{cylinder}
\end{figure}

Then, the supersonic flow past a cylinder~\cite{jiang_efficient_1996} is solved to examine the shock capturing capability of the FP schemes on the inhomogeneous wall-bounded curvilinear grid. The $M=2$ supersonic flow moves towards the cylinder and the slip wall boundary condition is imposed to the wall and supersonic inflow and outflow boundary condition are assigned to the left boundary and others, respectively. The grid is given by:
\begin{equation}
\begin{aligned}
 x&=\left( R_x-(R_x-1) {\eta}'  \right) cos\left( \theta(2\xi'-1)\right)\\
 y&=\left( R_y-(R_y-1) {\eta}'  \right) sin\left( \theta(2\xi'-1) \right)\\
  \xi'&=\dfrac{\xi-1}{i_{max}-1}, \xi=i+0.2\phi_i  \\
 \eta'&=\dfrac{\eta-1}{j_{max}-1}, \eta=j+0.2\sqrt{1-\phi_i^2}
\end{aligned}
\end{equation}
where $\theta=5\pi/12$, $R_x=3$, $R_y=6$ and $\phi_i$ is a random number distributed between $[0,1]$. The resolution of the grid is $i_{max}=61$ and $j_{max}=81$. The free stream pressure and density are $p=1$ and $\rho=\gamma$, respectively. The computational results with the global Lax-Friedrichs dissipation are evaluated after $t=25$. In Fig.~\ref{cylinder}, the pressure distributions of the original WENO schemes are significantly influenced by the non-physical oscillations, though in the smooth region away from the shock. However, the proposed FP schemes resolve the smooth region without noises and capture the shock waves sharply.
\subsubsection{\label{sec:WT}A Mach 3 wind tunnel with a step}
\begin{figure} [htb]
    \center
    \includegraphics[width=\textwidth]{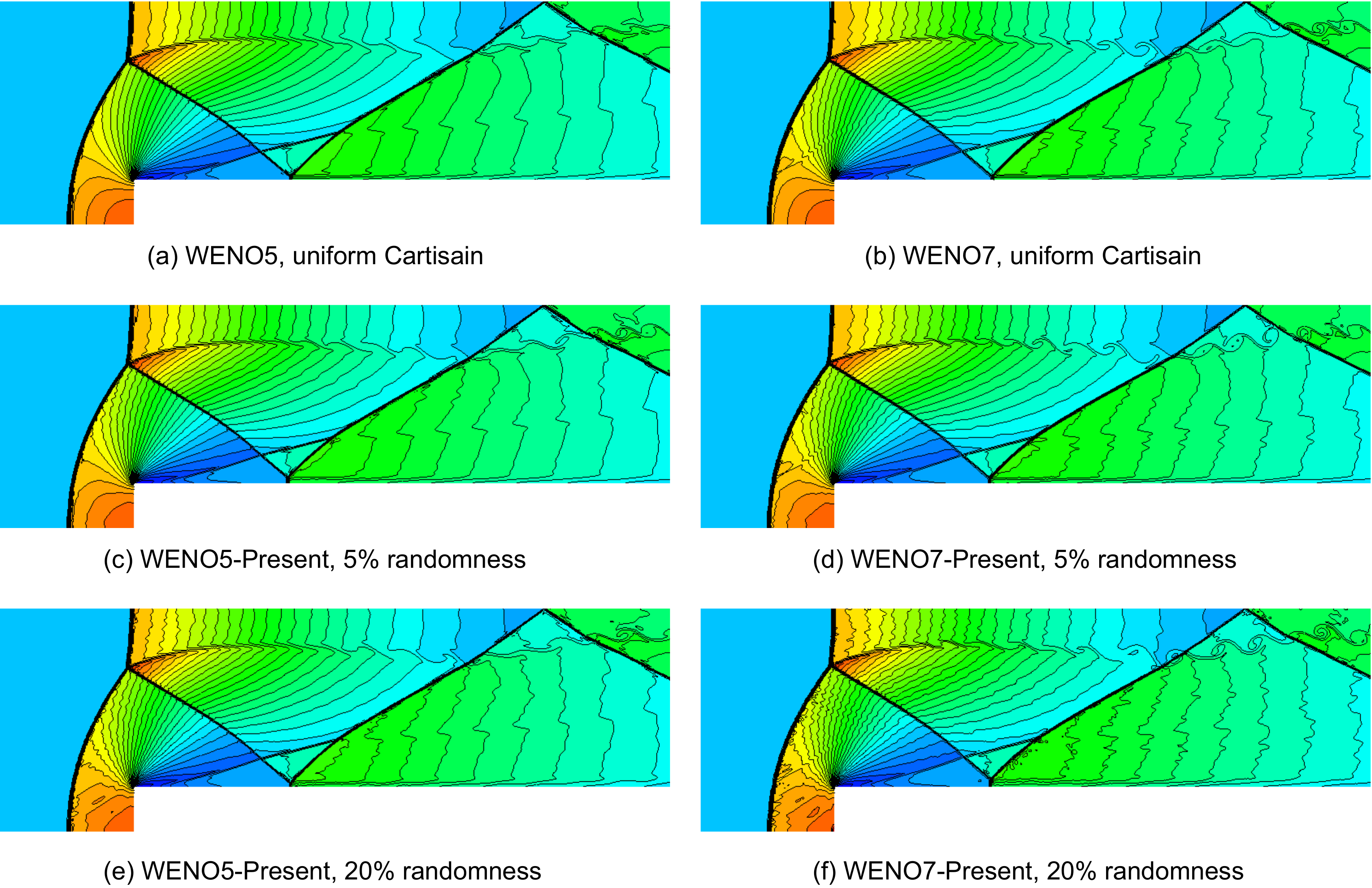}
    \caption{The density contours ranging from 0.2 to 6.5 of the flows in the wind tunnel with a step.}
    \label{fig:wind_step}
\end{figure}

Next, the simulation of the supersonic flow past the wind tunnel with a step is conducted to show the performance of the FP schemes on capturing the shock reflections and the high-resolution for shearing vortexes. 
The length and width of the wind tunnel are 3 and 1, respectively. 
The step in the bottom of the wind tunnel is located at 0.6 from the left boundary with a height of 0.2. The initial flow field is given by a right-going Mach 3 flow with $P=1$ and $\rho=\gamma=1.4$. 
The in-flow and out-flow boundary condition are applied to the left and right boundary.
And the reflective boundary conditions are considered along the walls of the tunnel. 
The computational domain is first discretized with a resolution of $\Delta x=1/200$ and then randomly disturbed by $5\%$ and $20\%$ of $\Delta x$, respectively.
In Fig.~\ref{fig:wind_step}c-\ref{fig:wind_step}f, the smooth density contours on the two severely disturbed grids at $t=4$ indicate that the present FP schemes indeed resolve this complex flow without spurious oscillations.
The well resolved reflective shocks around the wall of the wind tunnel and the vortexes generated in the shear layer give a good agreement with the results on the uniform grids.

\subsection{\label{viscous}Transonic aerodynamic problems}
\subsubsection{\label{sec:naca0012}Inviscid flow past a NACA0012 airfoil}
\begin{figure} [htbp]
    \center
    \includegraphics[width=\textwidth]{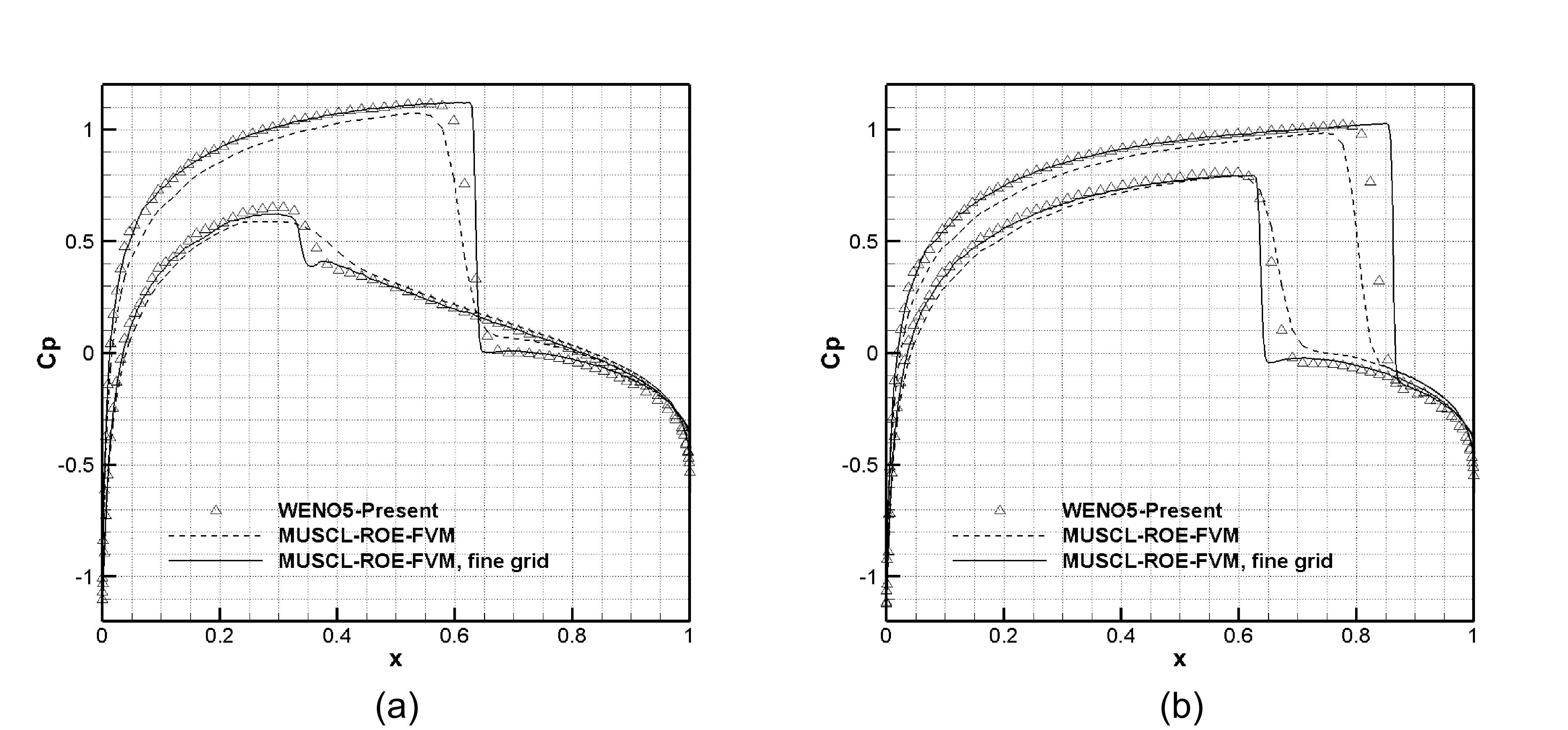}
    \caption{The pressure coefficient distributions along the wall of the NACA0012 airfoil. }
    \label{fig:naca0012_Cp}
\end{figure}

\begin{figure} [htbp]
    \center
    \includegraphics[width=\textwidth]{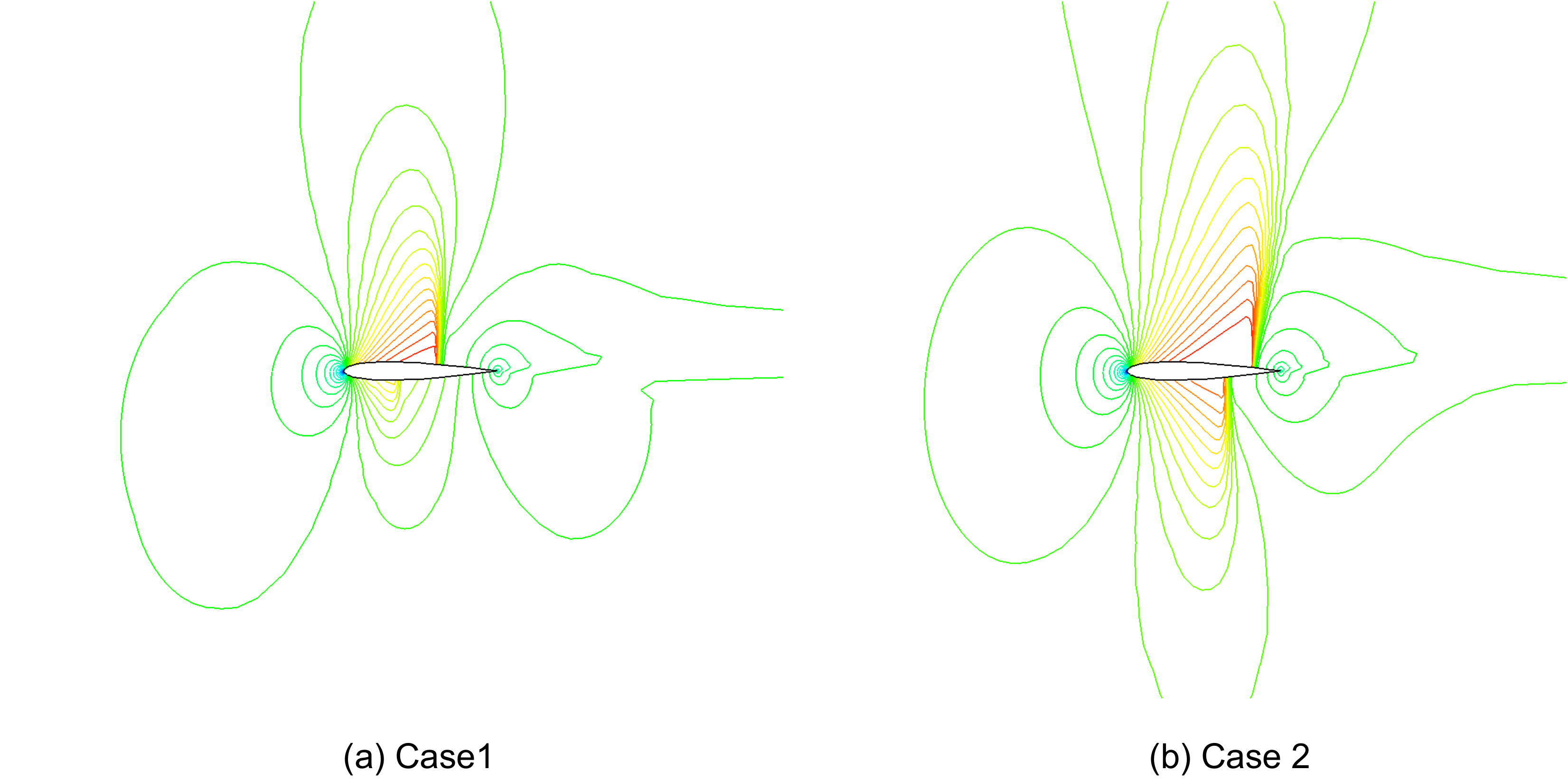}
    \caption{The Mach number contours from 0.172 to 1.325 of the transonic flow past the NACA0012 airfoil.}
    \label{fig:naca0012_Mach}
\end{figure}

In this case, we consider two inviscid transonic flows past a NACA0012 airfoil.
For the first case, Mach number $M=0.8$ and angle of attack $ AOA= 1.25^{\circ} $ (case 1).
And in the second case, Mach number $M=0.85$ and angle of attack $AOA= 1.0^{\circ}$ (case 2). 
A coarse O-type grid with $160 \times 32$ cells in circumferential and radial, respectively, as well as a fine grid with $1300 \times 180$ cells, is generated. 
Fig.~\ref{fig:naca0012_Cp} shows the profiles of pressure coefficients ($C_p$) of three simulations.
Two of them are conducted in the coarse grid and the fine grid, with the 2nd-order FVM of MUSCL reconstruction and ROE scheme (MUSCL-ROE-FVM).
The result in the fine grid is used as a reference data.
The other is performed in the coarse grid with the FP WENO5 scheme.
WENO-Present gives the sharper shock pattens and more accurate $C_p$ distributions, especially on the upper airfoil surface. 
Observed from the numerical Mach contours in Fig.~\ref{fig:naca0012_Mach}, the smooth and continuous flows on the curvilinear grid are obtained by the present FP scheme.
\subsubsection{\label{sec:M6}Transonic viscous flow pass the ONERA M6 wing}
\begin{figure} [htbp]
    \center
    \includegraphics[width=\textwidth]{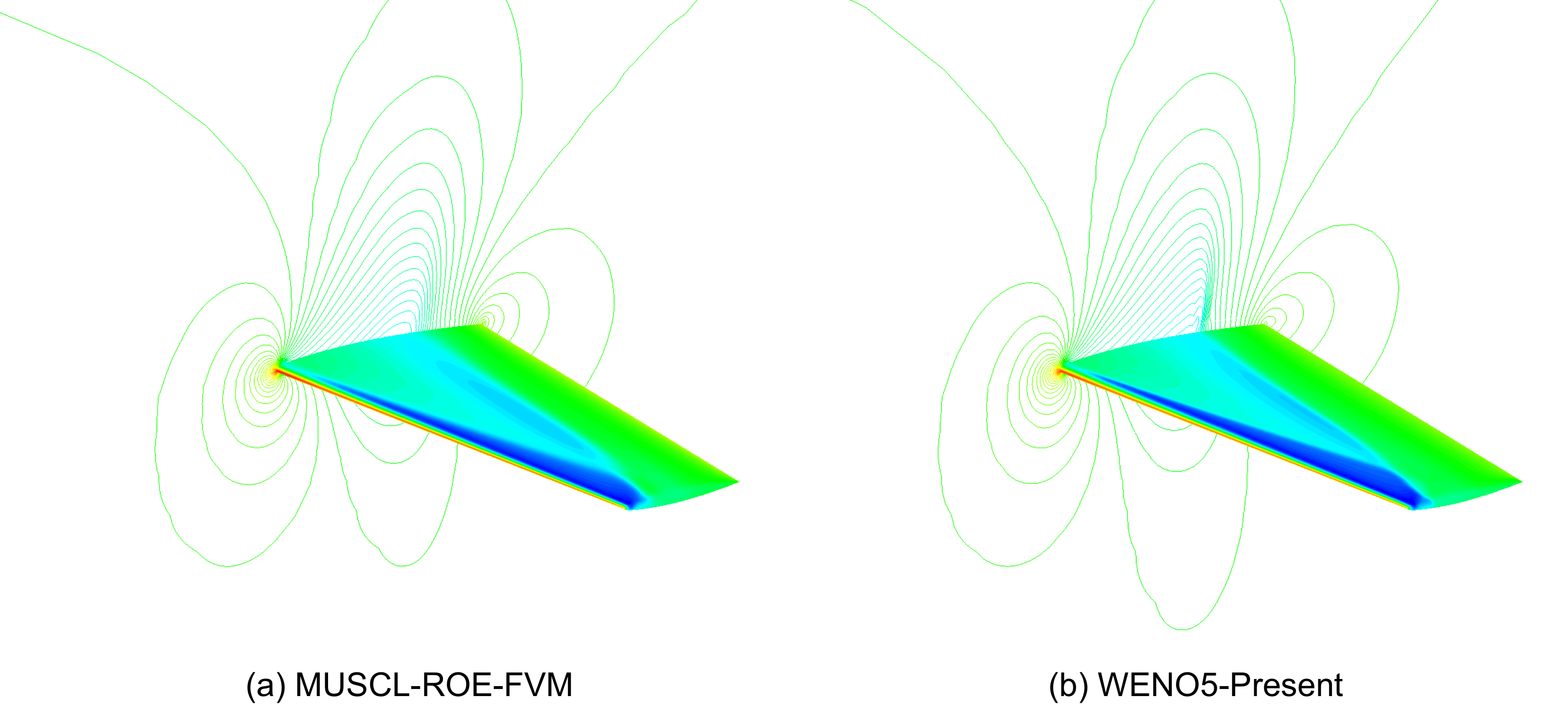}
    \caption{The pressure contours(from 130 Kpa to 490 Kpa) around the surface and the symmetry pane of the ONERA M6 wing.}
    \label{fig:M6_wing_contours}
\end{figure}

\begin{figure} [htbp]
    \center
    \includegraphics[width=0.9\textwidth]{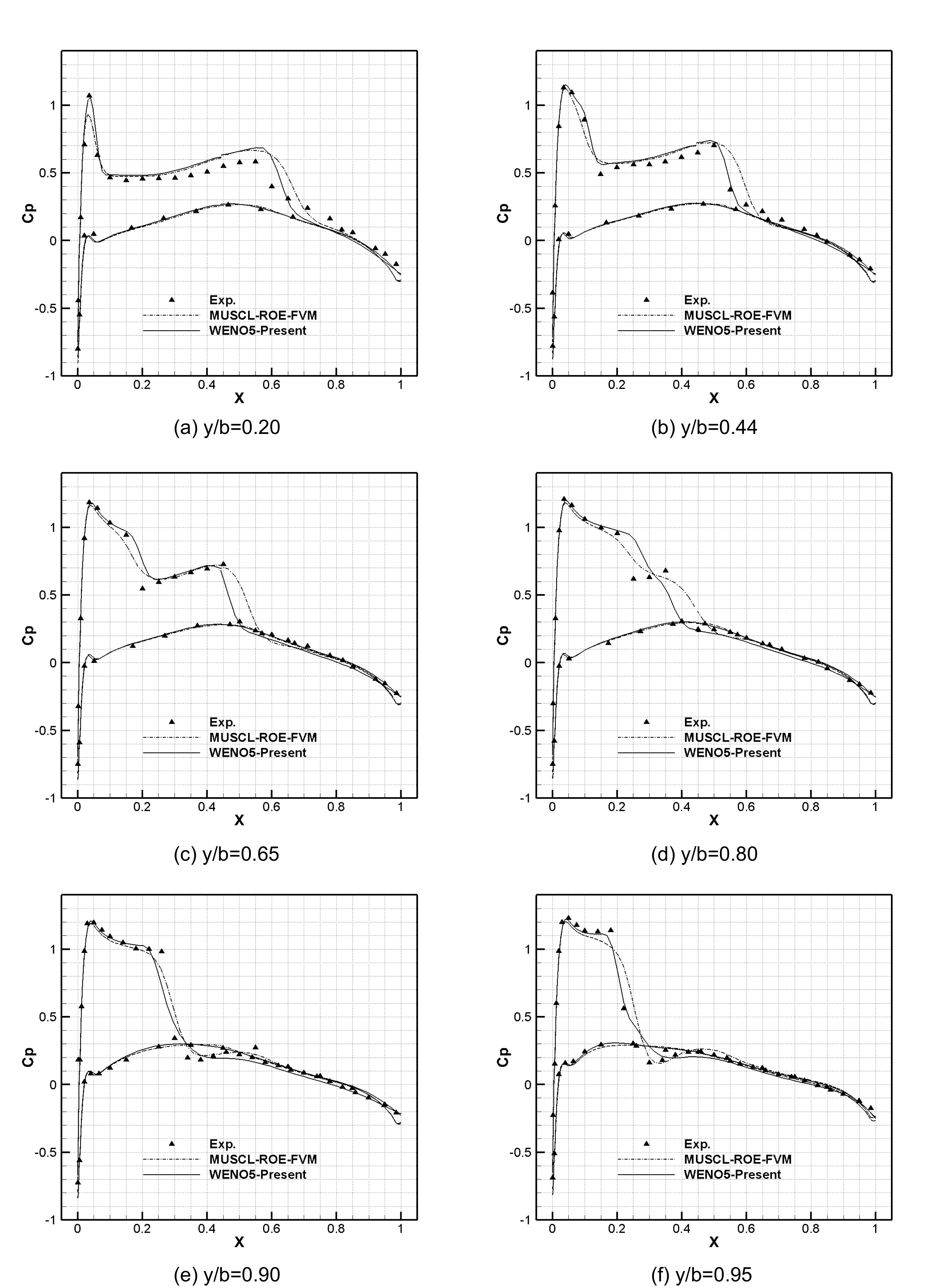}
    \caption{The pressure coefficient distributions at six spanwise cross-sections of ONERA M6 wing. }
    \label{fig:M6_wing_Cp}
\end{figure}

At last, a three-dimensional (3D) transonic flow pass the ONERA M6 wing is considered in this section. 
The simulation is conducted at a Mach number $M=0.84$ and an angle of attack $AOA=3.06^{\circ}$ with a Reynold number of $Re_l = 1.172\times10^7$ based on the mean aerodynamic chord of $l = 0.64607m$. 
The computational grid consists of 294912 cells and the simulation adopts the implicit time marching algorithm with Lower-Upper Symmetric-Gauss-Seidel (LUSGS) \cite{SEOKKWAN1988LUSGS} to reduce the computing costs. 
Here SA turbulence model~\cite{SA_model} is employed to close the viscous terms. 
The pressure contours illustrated in Fig.~\ref{fig:M6_wing_contours} indicate that WENO5-Present resolve the shock on the upper surface of M6 wing more sharply without ruining the smooth regions, in comparison to ROE-MUSCL-FVM. 
Moreover, $C_p$ profiles in Fig.~\ref{fig:M6_wing_Cp} show an improved agreement to the experimental data \cite{Schmitt1979PressureDO} by WENO5-Present, especially for shock capturing and the pressure peak near the leading edge on the upper surface.

\section{Concluding remarks}
In the present paper, we give a further investigation on the dissipation of WENO5 scheme by rewriting it to a combination of four sub-stencil reconstructions. 
Applying this treatment to three previous FP methods, we conclude and prove later that the FP identity can be maintained in WENO5 if the above four reconstructions of the metrics give an identical value. 
Then these results can be extended to other nonlinear schemes, such as WENOZ, WENO7, etc.
As a straightforward application of this condition, the new FP metrics are designed to replace the original one for nonlinear upwind schemes. With this, the original forms of these schemes can be maintained, which makes it convenient to carry out in different conservative finite difference schemes, such as WENO5, WENOZ and WENO7. 
After conducting an accuracy analysis of the central part and dissipation part of the proposed FP schemes, respectively, it is observed that the dissipation portion retains the original accuracy though the central part may be influenced by the new FP metrics. Therefore, a simple replacement with a high-order central part is suggested to compensate the accuracy loss. 
Some classical validations are conducted to demonstrate the accuracy and robustness of the present FP method,
%
%
whose results indicate that the present method indeed maintains the FP identity on the curvilinear grids with high accuracy and robust shock capturing ability.

\section*{Acknowledgements}
This work was supported by the National Natural Science Foundation of China (Grant Nos. 91952203 and 11902271) and 111 project on Aircraft Complex Flows and the Control (Grant No. B17037).

\section*{Appendix A. Accuracy analysis for the FP upwind schemes}  \label{sec:high_order}
\subsection*{The LU5 scheme}
The proposed FP metrics and Jacobian $g^*_{i-2}$, $g^*_{i-1}$, $g^*_{i+2}$ and $g^*_{i+3}$ are computed by the 3rd-order reconstruction from three 6th-order ones $g_{i}$, $g_{i+1/2}$ and $g_{i+1}$. Therefore, they maintain 3rd-order accuracy at least. In smooth regions, Taylor expansion of Eq.~\eqref{eq:FP_metrics} gives, respectively,
\begin{equation}\label{eq:metrcis_taylor}
\begin{aligned}
\Delta g_{i-2}^*=   g^*_{i-2}-g_{i-2}&=\dfrac{5}{2}\mathcal{R}'''\Delta \xi^3-\dfrac{7}{20}\mathcal{R}^{(4)}\Delta \xi^4+\dfrac{103}{240}\mathcal{R}^{(5)}\Delta \xi^5+O\left(\Delta \xi^6\right),\\
\Delta g_{i-1}^*=  g^*_{i-1}-g_{i-1}&=\dfrac{1}{2}\mathcal{R}'''\Delta \xi^3+\dfrac{1}{20}\mathcal{R}^{(4)}\Delta \xi^4+\dfrac{11}{240}\mathcal{R}^{(5)}\Delta \xi^5+O\left(\Delta \xi^6\right),\\
\Delta g_{i}^*=  g^*_{i}-g_{i}&=0,\\
\Delta g_{i+1}^*=  g^*_{i+1}-g_{i+1}&=0,\\
\Delta g_{i+2}^*=  g^*_{i+2}-g_{i+2}&=-\dfrac{1}{2}\mathcal{R}'''\Delta \xi^3-\dfrac{9}{20}\mathcal{R}^{(4)}\Delta \xi^4-\dfrac{59}{240}\mathcal{R}^{(5)}\Delta \xi^5+O\left(\Delta \xi^6\right),\\
\Delta g_{i+3}^*=  g^*_{i+3}-g_{i+3}&=-\dfrac{5}{2}\mathcal{R}'''\Delta \xi^3-\dfrac{57}{20}\mathcal{R}^{(4)}\Delta \xi^4-\dfrac{487}{240}\mathcal{R}^{(5)}\Delta \xi^5+O\left(\Delta \xi^6\right),\\
\end{aligned}
\end{equation}
where $\mathcal{R}'''=\mathcal{R}'''(\xi)$ and $\mathcal{R}^{(4)}=\mathcal{R}^{(4)}(\xi)$ are the third and fourth derivatives at $i$ of the primary function $\mathcal{R}\left(\xi\right)$,
\begin{equation}
g_i=\dfrac{1}{\Delta \xi}\int_{i-1/2}^{i+1/2} \mathcal{R}\left(\xi\right) d\xi.
\end{equation}

If the FP metrics and Jacobian are adopted to the dissipation part of the LU5 scheme, we can obtain the FP conservative variables for the dissipation 
\begin{equation} \label{eq:linear_order1}
\begin{aligned}
\boldsymbol{\tilde{Q}}_{m}^{*}&=\boldsymbol{\tilde{Q}}_{m}+\left( \boldsymbol{\tilde{Q}}_{m}^{*}-\boldsymbol{\tilde{Q}}_{m} \right) ,m=i-2,i+3\\
&=\boldsymbol{\tilde{Q}}_{m}+\boldsymbol{Q}_{m}\Delta g_m^{*},
\end{aligned}
\end{equation}
and the dissipation at cell-face $i+1/2$ is
\begin{equation} \label{eq:linear_order2}
\begin{aligned}
 \boldsymbol{D}_{i+1/2}=&\left( \boldsymbol{\tilde{Q}}_{i-2}^{*}-5\boldsymbol{\tilde{Q}}_{i-1}^{*}+10\boldsymbol{\tilde{Q}}_{i}^{*}-10\boldsymbol{\tilde{Q}}_{i+1}^{*}+5\boldsymbol{\tilde{Q}}_{i+2}^{*}-\boldsymbol{\tilde{Q}}_{i+3}^{*} \right)\\
 =&\left( \boldsymbol{\tilde{Q}}_{i-2}-5\boldsymbol{\tilde{Q}}_{i-1}+10\boldsymbol{\tilde{Q}}_{i}-10\boldsymbol{\tilde{Q}}_{i+1}+5\boldsymbol{\tilde{Q}}_{i+2}-\boldsymbol{\tilde{Q}}_{i+3} \right)\\
 &+\left( 10\mathcal{R}'''\boldsymbol{Q}_{i+1/2}''+5\mathcal{R}^{(4)}\boldsymbol{Q}_{i+1/2}'+\mathcal{R}^{(5)}\boldsymbol{Q}_{i+1/2} \right)\Delta \xi^5+\boldsymbol{O}\left( \Delta \xi^6 \right),
\end{aligned}
\end{equation}
where $g$ refers to the Jacobian $1/J$. Similarly, we have
\begin{equation} \label{eq:linear_order3}
\begin{aligned}
 \boldsymbol{D}_{i-1/2}=&\left( \boldsymbol{\tilde{Q}}_{i-3}-5\boldsymbol{\tilde{Q}}_{i-2}+10\boldsymbol{\tilde{Q}}_{i-1}-10\boldsymbol{\tilde{Q}}_{i}+5\boldsymbol{\tilde{Q}}_{i+1}-\boldsymbol{\tilde{Q}}_{i+2} \right)\\
 &+\left( 10\mathcal{R}'''\boldsymbol{Q}_{i-1/2}''+5\mathcal{R}^{(4)}\boldsymbol{Q}_{i-1/2}'+\mathcal{R}^{(5)}\boldsymbol{Q}_{i-1/2} \right)\Delta \xi^5+\boldsymbol{O}\left( \Delta \xi^6 \right).
\end{aligned}
\end{equation}
The additional terms retain 5th-order accuracy in the conservative FDM because
\begin{equation} \label{eq:linear_order4}
\begin{aligned}
\dfrac{ \boldsymbol{D}_{i+1/2}-\boldsymbol{D}_{i-1/2} }{\Delta \xi}=&\left[ \left( \boldsymbol{\tilde{Q}}^{(5)}_{i+1/2} - \boldsymbol{\tilde{Q}}^{(5)}_{i-1/2}\right)+ 10\mathcal{R}'''\left(\boldsymbol{Q}_{i+1/2}''-\boldsymbol{Q}_{i-1/2}''\right) \right. \\
& \left. +5\mathcal{R}^{(4)}\left(\boldsymbol{Q}_{i+1/2}'-\boldsymbol{Q}_{i-1/2}'\right)+\mathcal{R}^{(5)}\left(\boldsymbol{Q}_{i+1/2}-\boldsymbol{Q}_{i-1/2}\right) \right] \Delta \xi^4+\boldsymbol{O}\left( \Delta \xi^5 \right)\\
& =\left( \boldsymbol{\tilde{Q}}^{(6)}+10\mathcal{R}'''\boldsymbol{Q}''' +5\mathcal{R}^{(4)} \boldsymbol{Q}''+\mathcal{R}^{(5)}\boldsymbol{Q}' \right) \Delta \xi^5 +\boldsymbol{O}\left( \Delta \xi^5 \right)
\end{aligned}
\end{equation}
Obviously, although the proposed 3rd-order metrics and Jacobian are applied to the dissipation of the LU5 scheme, only the extra $\boldsymbol{O}\left( \Delta \xi^5 \right)$ terms are added to the standard terms such that the analytic convergence order of the proposed upwind dissipation still maintains 5th-order accuracy.

\subsection*{The WENO5 scheme}
According to Ref.~\cite{jiang_efficient_1996}, the WENO5 scheme is a convex combination of the 3rd-order reconstruction of all the candidate sub-stencils
\begin{equation}\label{eq:WENO-Flux}
\begin{aligned}
  \tilde{f}_{i+1/2}=q^{5}(\tilde{f}_{i-2},\cdots,\tilde{f}_{i+2})+\sum\limits_{k=0}^{2}\left( \omega_k-C_k \right)q^3_k \left( \tilde{f}_{i+k-2},\tilde{f}_{i+k-1}, \tilde{f}_{i+k}\right).
\end{aligned}
\end{equation}
As given by Borges et al.~\cite{borges_improved_2008}, the conservative FDM maintains the 5th-order accuracy if the nonlinear weights $\omega_k$ and $q_k$ in sub-stencils satisfy the followings,
\begin{align}
\label{eq:ww1} &\omega_k =C_k + O(\Delta \xi^2),\\
\label{eq:ww2} &q_k=\tilde{h}_{i+1/2}+O(\Delta \xi^3),\\
\label{eq:ww3} &\sum\limits_{k=0}^{2} A_k\left(\omega_k-\omega_k^{'}\right)=O(\Delta \xi^3),
\end{align}
 where $\omega_k^{'}$ denotes the nonlinear weight for $\tilde{f}_{i-1/2}$. It should be noted that Eq.~\eqref{eq:ww3} is ignored in the standard WENO5 scheme of Jiang and Shu~\cite{jiang_efficient_1996} due to the large $\epsilon=1\times 10^{-6}$.

If the proposed metrics and Jacobian $g^*_{m}$ and the 6th-order one $g_{m}$ are adopted to calculate cell-averaged fluxes and apply the fluxes splitting, denoted by $\boldsymbol {\tilde{F}}^{*+}$ and $\boldsymbol {\tilde{F}}^{+}$, respectively, we can obtain the relation between them as
\begin{equation}
\begin{aligned}
\boldsymbol {\tilde{F}}^{*+}=&\boldsymbol {\tilde{F}}^{+}+\boldsymbol {L} \left\lbrace  \boldsymbol{F}\left[\left(\dfrac{\xi_x}{J}\right)^{*}-\left(\dfrac{\xi_x}{J}\right)\right]
+\boldsymbol{G}\left[\left(\dfrac{\xi_y}{J}\right)^{*}-\left(\dfrac{\xi_y}{J}\right)\right] \right.\\
&\qquad \left. +\boldsymbol{H}\left[\left(\dfrac{\xi_z}{J}\right)^{*}-\left(\dfrac{\xi_z}{J}\right)\right]
+\boldsymbol{\lambda} \boldsymbol{Q}\left[\left(\dfrac{1}{J}\right)^{*}-\left(\dfrac{1}{J}\right)\right] \right\rbrace\\
&=\boldsymbol {\tilde{F}}^{+}+\boldsymbol {O}(\Delta \xi^3) .
\end{aligned}
\end{equation}
 Therefore, the reconstructed cell-face fluxes $q_k^{*+}$ in the sub-stencil by $\tilde{f}^{*+}_{m}$ which is one of the component of $\boldsymbol {\tilde{F}}_m^{*+}$ can still maintain 3rd-order accuracy because of Eq.~\eqref{eq:metrcis_taylor}. To simplify notation, we drop the superscript $+$ for $\tilde{f}^{*+}$, $q^{*+}$ and $\beta^{*+}$ in the followings. In details,
\begin{equation}
\begin{aligned} \label{eq:sub1_condition}
q_0^{*}&= \dfrac{1}{3}\tilde{f^*}_{i-2}
          -\dfrac{7}{6}\tilde{f^*}_{i-1}
          +\dfrac{11}{6}\tilde{f^*}_{i}  \\
       &= \left(\dfrac{1}{3}\tilde{f}_{i-2}
          -\dfrac{7}{6}\tilde{f}_{i-1}
          +\dfrac{11}{6}\tilde{f}_{i}\right) +  O(\Delta \xi^3) \\
       &= \tilde{h}_{i+1/2}+O(\Delta \xi^3)
\end{aligned}
\end{equation}
Similarly,
\begin{equation} \label{eq:sub2_condition}
q_1^{*}= \tilde{h}_{i+1/2}+O(\Delta \xi^3)
\end{equation}
\begin{equation} \label{eq:sub3_condition}
q_2^{*}= \tilde{h}_{i+1/2}+O(\Delta \xi^3).
\end{equation}

To investigate the accuracy of the nonlinear weights, we define
\begin{equation}
\begin{aligned}
\tilde{f^*}_{m}&= \tilde{f}_{m}+\left(\tilde{f^*}_{m}-\tilde{f}_{m}\right) \\
                   &= \tilde{f}_{m}+f_m \Delta g_m^* , m=i-2,\cdots,i+2,
\end{aligned}
\end{equation}
and
\begin{equation}
\begin{aligned}
 f_m \Delta g_m^*=\boldsymbol{l}_s\left[\boldsymbol{F}_m \Delta\left( \dfrac{\xi_x}{J} \right)^* + \boldsymbol{G}_m \Delta\left( \dfrac{\xi_y}{J} \right)^* + \boldsymbol{H}_m \Delta\left( \dfrac{\xi_z}{J} \right)^* + \lambda_s \boldsymbol{Q}_m \Delta\left( \dfrac{1}{J} \right)^*  \right],
\end{aligned}
\end{equation}
which represents for the difference between $\tilde{f^*}_{m}$ and $\tilde{f}_{m}$. Then, the smoothness indicators can be given by
\begin{equation} \label{eq:IS0_star}
\begin{aligned}
\beta_0^{*}=&\quad \dfrac{1}{4} \left[  \left(\tilde{f}_{i-2}-4\tilde{f}_{i-1}+3\tilde{f}_{i} \right)
    +\left( f_{i-2}\Delta g_{i-2}^*-4f_{i-1}\Delta g_{i-1}^*+3f_{i} \Delta g_{i}^* \right) \right]^2\\
            &+\dfrac{13}{12}\left[  \left( \tilde{f}_{i-2}-2\tilde{f}_{i-1}+\tilde{f}_{i} \right)
    + \left( f_{i-2}\Delta g_{i-2}^*-2f_{i-1}\Delta g_{i-1}^*+f_{i}\Delta g_{i}^*\right) \right]^2,
\end{aligned}
\end{equation}
\begin{equation} \label{eq:IS1_star}
\begin{aligned}
\beta_1^{*}=&\quad \dfrac{1}{4} \left[  \left(\tilde{f}_{i-1}-\tilde{f}_{i+1} \right)
    +\left( f_{i-1}\Delta g_{i-1}^*-f_{i+1}\Delta g_{i+1}^* \right) \right]^2\\
            &+\dfrac{13}{12}\left[  \left( \tilde{f}_{i-1}-2\tilde{f}_{i}+\tilde{f}_{i+1} \right)
    + \left( f_{i-1}\Delta g_{i-1}^*-2f_{i}\Delta g_{i}^*+f_{i+1}\Delta g_{i+1}^*\right) \right]^2,
\end{aligned}
\end{equation}
\begin{equation} \label{eq:IS2_star}
\begin{aligned}
\beta_2^{*}=&\quad \dfrac{1}{4} \left[  \left(3\tilde{f}_{i}-4\tilde{f}_{i+1}+\tilde{f}_{i+2} \right)
    +\left( 3f_{i}\Delta g_{i}^*-4f_{i+1}\Delta g_{i+1}^*+f_{i+2}\Delta g_{i+2}^* \right) \right]^2\\
            &+\dfrac{13}{12}\left[  \left( \tilde{f}_{i}-2\tilde{f}_{i+1}+\tilde{f}_{i+2} \right)
    + \left( f_{i}\Delta g_{i}^*-2f_{i+1}\Delta g_{i+1}^*+f_{i+2}\Delta g_{i+2}^*\right) \right]^2.
\end{aligned}
\end{equation}

In smooth regions, Taylor expansion of Eq.~\eqref{eq:IS0_star} at $i$ gives,
\begin{equation} \label{eq:IS0_taylor}
\begin{aligned}
\beta_0^{*}=&\tilde{f}'^2\Delta \xi^2+\left(\dfrac{13}{12}\tilde{f}''^2-\dfrac{2}{3}\tilde{f}'\tilde{f}''' +\dfrac{1}{2}\tilde{f}'\mathcal{R}'''f\right)\Delta \xi^4\\
&-\left( \dfrac{13}{6}\tilde{f}''\tilde{f}'''-\dfrac{1}{2}\tilde{f}'\tilde{f}^{(4)}-\dfrac{13}{4}\tilde{f}''\mathcal{R}'''f+3\tilde{f}'\mathcal{R}'''f'+\dfrac{11}{20}\tilde{f}'\mathcal{R}^{(4)}f\right)\Delta \xi^5+O(\Delta \xi^6),
\end{aligned}
\end{equation}
\begin{equation} \label{eq:IS1_taylor}
\begin{aligned}
\beta_1^{*}=&\tilde{f}'^2\Delta \xi^2+\left(\dfrac{13}{12}\tilde{f}''^2+\dfrac{1}{3}\tilde{f}'\tilde{f}''' -\dfrac{1}{2} \tilde{f}'\mathcal{R}'''f\right)\Delta \xi^4\\
&+\left( \dfrac{13}{12}\tilde{f}''\mathcal{R}'''f+\dfrac{1}{2}\tilde{f}'\mathcal{R}'''f'-\dfrac{1}{20}\tilde{f}'\mathcal{R}^{(4)}f\right)\Delta \xi^5+O(\Delta \xi^6),\qquad     \qquad     \qquad     \qquad     \quad
\end{aligned}
\end{equation}
\begin{equation} \label{eq:IS2_taylor}
\begin{aligned}
\beta_2^{*}=&\tilde{f}'^2\Delta \xi^2+\left(\dfrac{13}{12}\tilde{f}''^2-\dfrac{2}{3}\tilde{f}'\tilde{f}''' +\dfrac{1}{2} \tilde{f}'\mathcal{R}'''f\right)\Delta \xi^4\\
&+\left( \dfrac{13}{6}\tilde{f}''\tilde{f}'''-\dfrac{1}{2}\tilde{f}'\tilde{f}^{(4)}-\dfrac{13}{12}\tilde{f}''\mathcal{R}'''f+\tilde{f}'\mathcal{R}'''f'+\dfrac{9}{20}\tilde{f}'\mathcal{R}^{(4)}f\right)\Delta \xi^5+O(\Delta \xi^6).
\end{aligned}
\end{equation}

Therefore, applying the 3rd-order metrics and Jacobian to calculate the smoothness indicators $\beta_k^{*}$ does not violate the convergence orders of them. Furthermore,
it is straightforward to see that

 (1) for the WENO5-Present scheme, we obtain
\begin{equation}\label{eq:worderweno-js}
\beta_k^{*}=
\begin{cases}
   \left(\tilde{f}'\Delta \xi \right)^2 \left( 1+O(\Delta \xi^2) \right) &\tilde{f}'\neq0\\
   \dfrac{13}{12}\left(\tilde{f}''\Delta \xi^2 \right)^2 \left( 1+O(\Delta \xi) \right) &\tilde{f}'=0,
\end{cases}
\end{equation}
with $k=0,1,2$, which is the same with the original WENO5 scheme proposed by Jiang and Shu~\cite{jiang_efficient_1996};

(2) for WENOZ-Present scheme, we obtain
\begin{equation}\label{eq:worderweno-z}
\begin{aligned}
\tau_5^*&=|\beta_0^{*}-\beta_2^{*}|
&=\left( \dfrac{13}{3}\tilde{f}''\tilde{f}'''-\tilde{f}'\tilde{f}^{(4)}-\dfrac{13}{3}\tilde{f}''\mathcal{R}'''f_i+4\tilde{f}'\mathcal{R}'''f'+\tilde{f}'\mathcal{R}^{(4)}f\right)\Delta \xi^5+O(\Delta \xi^6),
\end{aligned}
\end{equation}
\begin{equation}\label{eq:worderweno-z-weights}
\begin{aligned}
\left( 1+\dfrac{\tau_5^*}{\beta_k^*} \right)=1+O(\Delta \xi^3)
\end{aligned}
\end{equation}
 whose truncation error is the same order with the original WENOZ scheme suggested by Borges et al.~\cite{borges_improved_2008}.
Then, Eqs.~\eqref{eq:worderweno-js} $\sim$ \eqref{eq:worderweno-z-weights} indicate that the orders of the nonlinear weights are retained by applying the present FP metrics and Jacobian.

Finally, we conclude that the sufficient conditions given in Eqs.~\eqref{eq:ww1} $\sim$ \eqref{eq:ww3} are all satisfied in the present FP schemes. Therefore, the dissipation parts of the present WENO5 retain 5th-order accuracy at non-critical points as the same as the original WENO5 and WENOZ scheme.

\section*{Appendix B. Extension to WENO7}  \label{sec:WENO7_metrics}
For the WENO7 scheme, the FP metrics, represented as $g^*_{1+1/2}$, $g^*_{i-3}$, $\cdots$, $g^*_{i+4}$, can be defined as
\begin{equation}\label{eq:FP_metrics_weno7}
\left \{
\begin{aligned}
g^*_{i-1}&=g_{i-1}, \ g^*_{i}=g_{i},  \\
g^*_{i+1}&=g_{i+1}, \ g^*_{i+2}=g_{i+2}, \\
\dfrac{3}{12}g^*_{i-3}&=\dfrac{13}{12}g^*_{i-2}-\dfrac{23}{12}g^*_{i-1}+\dfrac{25}{12}g^*_{i}-g^*_{i+1/2}, \\
\dfrac{1}{12}g^*_{i-2}&=\dfrac{5}{12}g^*_{i-1}-\dfrac{13}{12}g^*_{i}-\dfrac{3}{12}g^*_{i+1}+g^*_{i+1/2}, \\
\dfrac{1}{12}g^*_{i+3}&=-\dfrac{3}{12}g^*_{i+1}-\dfrac{13}{12}g^*_{i+2}+\dfrac{5}{12}g^*_{i+3}+g^*_{i+1/2} ,\\
\dfrac{3}{12}g^*_{i+4}&=\dfrac{25}{12}g^*_{i+2}-\dfrac{13}{12}g^*_{i+3}+\dfrac{13}{12}g^*_{i+4}-g^*_{i+1/2} .
\end{aligned}
 \right.
\end{equation}
where
 \begin{equation}\label{eq:half_metrics8}
g^*_{i+1/2}=\dfrac{1}{12}\left( -g_{i-1}+7g_{i}+7g_{i+1}-g_{i+2} \right).
\end{equation}

The FP identity and the accuracy of the central part is achieved by adding an extra term, which can compensate the loss of the accuracy due to the use of the 4th-order metrics. In details,
\begin{equation}\label{Flux_cor}
\begin{aligned}
\boldsymbol{\tilde{F}}^{C4}_{i+1/2}&=\dfrac{1}{840}\left( -3\boldsymbol{\tilde{F}}^*_{i-3}+29\boldsymbol{\tilde{F}}^*_{i-2}-139\boldsymbol{\tilde{F}}^*_{i-1}+533\boldsymbol{\tilde{F}}^*_{i}+533\boldsymbol{\tilde{F}}^*_{i+1}-139\boldsymbol{\tilde{F}}^*_{i+2}+29\boldsymbol{\tilde{F}}^*_{i+3}-3\boldsymbol{\tilde{F}}^*_{i+4} \right),\\
\boldsymbol{\tilde{F}}^{C8}_{i+1/2}&=\dfrac{1}{840}\left( -3\boldsymbol{\tilde{F}}_{i-3}+29\boldsymbol{\tilde{F}}_{i-2}-139\boldsymbol{\tilde{F}}_{i-1}+533\boldsymbol{\tilde{F}}_{i}+533\boldsymbol{\tilde{F}}_{i+1}-139\boldsymbol{\tilde{F}}_{i+2}+29\boldsymbol{\tilde{F}}_{i+3}-3\boldsymbol{\tilde{F}}_{i+4} \right),\\
\boldsymbol{\tilde{F}}_{i+1/2}^{'}&=\boldsymbol{\tilde{F}}^*_{i+1/2}+\left(\boldsymbol{\tilde{F}}^{C8}_{i+1/2}-\boldsymbol{\tilde{F}}^{C4}_{i+1/2}\right),
\end{aligned}
\end{equation}
where $\boldsymbol{\tilde{F}}^*_{m}$ and $\boldsymbol{\tilde{F}}_{m}$ are the cell-averaged fluxes calculated by the 4th- and 8th-order metrics, respectively. Finally, the new cell-face fluxes $\boldsymbol{\tilde{F}}_{i+1/2}^{'}$ can obtain the high-order accuracy and maintain the FP identity at the same time.

\section*{References}

\bibliographystyle{elsarticle-num}
\bibliography{main}

\end{document}